%% file: cenarro_rev2.tex
\documentstyle[psfig,graphicx]{mn}                       

\title[Mg and TiO spectral features at the near-IR] {Mg and TiO
  spectral features at the near-IR: Spectrophotometric index
  definitions and empirical calibrations}

\author[A.J. Cenarro et al.]  
    {A.J.~Cenarro$^1$,\thanks{E-mail: cenarro@iac.es}
    N.~Cardiel$^2$, A.~Vazdekis$^1$, and J.~Gorgas$^2$ 
\\ $^1$Instituto de Astrof\'{\i}sica de Canarias,
    E-38200, La Laguna, Tenerife, Spain.\\$^2$Depto. de Astrof\'{\i}sica, Fac. de
    Ciencias F\'{\i}sicas, Universidad Complutense de Madrid, E-28040
    Madrid, Spain.\\
}
\date{Accepted 2009 March 26. Received 2008 December 4}
\pagerange{\pageref{firstpage}--\pageref{lastpage}}
\pubyear{2000}

\def\LaTeX{L\kern-.36em\raise.3ex\hbox{a}\kern-.15em
    T\kern-.1667em\lower.7ex\hbox{E}\kern-.125emX}

\begin{document}

\label{firstpage}

\maketitle

\begin{abstract}

Using the near-infrared spectral stellar library of Cenarro et
al.~(2001a,b), the behaviours of the Mg\,{\sc i} line at 8807~\AA\ and
nearby TiO bands are analyzed in terms of the effective temperature,
surface gravity, and metallicity of the library stars. New
spectroscopic indices for both spectral features ---namely MgI and
sTiO--- are defined, and their sensitivities to different
signal-to-noise ratios, spectral resolutions, flux calibrations, and
sky emission line residuals are characterized.  The new two indices
exhibit interesting properties. In particular, MgI reveals as a good
indicator of the Mg abundance, whereas sTiO is a powerful
dwarf-to-giant discriminator for cold spectral types. Empirical
fitting polynomials that reproduce the strength of the new indices as
a function of the stellar atmospheric parameters are computed, and a
{\sc fortran} routine with the fitting functions predictions is made
available.  A thorough study of several error sources, non-solar
[Mg/Fe] ratios, and their influence on the fitting function residuals
is also presented. From this analysis, a [Mg/Fe] underabundance of
$\sim -0.04$ is derived for the Galactic open cluster M67.
\end{abstract}


\section{Introduction}

A powerful approach to unravel the stellar content of unresolved
stellar systems is to interpret the integrated strengths of key
spectral features on the basis of evolutionary stellar population
synthesis models (e.g. Worthey 1994; Vazdekis et al.~1996; Vazdekis
1999; Vazdekis et al.~2003, hereafter VAZ03; Thomas, Maraston \&
Bender 2003; Bruzual \& Charlot 2003; Maraston 2005; Schiavon
2007). These models make use of theoretical isochrones and spectral
stellar libraries to predict integrated line-strengths and/or spectral
energy distributions (SEDs) corresponding to simple stellar
populations (SSPs) of a given age, overall metallicity, abundance
pattern, initial mass function, and star formation history.

Up to date, major progress in this kind of studies has been achieved
in the optical spectral range. The Lick/IDS stellar library (Gorgas et
al.~1993; Worthey et al.~1994) has constituted so far the reference
system for most optical work on this topic. However, thanks to the
developing of much improved optical stellar libraries that superseed
the capabilities of the Lick/IDS library, like e.g.~Jones (1998),
ELODIE (Prugniel \& Soubiran 2001, 2004; Prugniel et al.~2007), STELIB
(Le Borgne et al.~2003), the Indo-US stellar library (Vald\'es et
al.~2004), and MILES (S\'anchez-Bl\'azquez et al.~2006; Cenarro et
al.~2007), a new generation of SSP models in the optical region
(e.g.~Vazdekis 1999; Bruzual \& Charlot 2003; P\'EGASE-HR, by Le
Borgne et al.~2004; Vazdekis et al.~2009, in preparation) is now
available. The larger spectral coverage and better spectral resolution
of the new models have motivated the development of new analysis
approaches that, rather than focusing on single spectral features, are
based on fitting techniques over the full spectrum that are
potentially useful for reconstructing in first order the star
formation histories of galaxies (e.g. Panter et al.~2003; Cid
Fernandes et al.~2005; Mathis, Charlot, \& Brinchmann~2006; Ocvirk et
al~2006a,b; Panter, Heavens, \& Jimenez 2007; Koleva et al.~2008).

Aside from the above work, there exists an important effort to advance
in our understanding of complementary spectral regions which are
governed by different types of stars, like the ultraviolet
(e.g.~Fanelli et al~1992; Gregg et al.~2004; Heap \& Lindler 2007) and
the infrared (e.g.~Ivanov et al.~2004; Ranade et al.~2004, 2007a,b;
M\'armol-Queralt\'o et al.~2008). In particular, aimed at providing
reliable SSP model predictions for the near-infrared (near-IR)
spectral region around Ca\,{\sc ii} triplet at $\sim 8600$~\AA, an
extensive spectral stellar library at $\lambda\lambda~8348-9020$~\AA\
(FWHM$ = 1.5$~\AA) that comprises 706 stars over a wide range of
atmospheric parameters was developed by Cenarro et al.~(2001a;
hereafter CEN01a). Subsequent libraries like STELIB and the Indo-US
also include the Ca\,{\sc ii} triplet region.  Initially, the library
in Cenarro et al.~(2001) was particularly devoted to understand the
behaviour of the Ca\,{\sc ii} triplet in individual stars. With this
aim, improved line-strength indices for this spectral feature (namely
CaT, PaT, CaT$^{*}$) which are especially suited to be measured in the
integrated spectra of stellar populations were defined (CEN01a). Also,
to minimize uncertainties and systematic errors of the empirical
calibration of the indices, a homogeneous system of revised
atmospheric parameters for the library stars was derived in Cenarro et
al.~2001b (hereafter CEN01b). Puting all these ingredients together,
the behaviour of the Ca\,{\sc ii} indices as a function of the stellar
atmospheric parameters was computed by means of so-called empirical
fitting functions (Cenarro et al.~2002; hereafter CEN02), which were
implemented into the evolutionary synthesis code of VAZ03 to predict
the integrated indices and the near-IR SEDs for SSPs of different
ages, metallicities, and IMFs.

The present paper can be considered as an extension of the above
project to two nearby spectral features: the Mg\,{\sc i} line at
$\lambda$~8807~\AA\ and the molecular bands of TiO at
$\lambda\lambda$~8432, 8442, 8452~\AA\ and $\lambda\lambda$~8860,
8868~\AA. As it was already shown in Cenarro et al.~(2003) for a
sample or 35 early-type galaxies, both spectral features ---together
with the Ca\,{\sc ii} triplet--- can play an important role to
characterize the properties of old and intermediate-aged stellar
populations. The fact that Mg is overabundant with respect to Fe in
massive elliptical galaxies and tightly correlates with the velocity
dispersion (e.g.~Dressler et al.~1987; Worthey et al.~1992; and
others) turns Mg indices into a key element to constrain galaxy
star-formation and evolution theories. Also, it is worth stressing the
importance of calibrating several spectroscopic indicators in a
relatively narrow spectral region, as in contrast with stellar
population studies in which blue and red indicators are employed
together, the ages and metallicities derived from a set of nearby
indices are expected to be consistent even for composite stellar
populations.

Therefore, the main objective of this paper is to carry out a
comprehensive study of the Mg\,{\sc i} and TiO features in individual
stars, so that their dependences with the atmospheric parameters are
calibrated and quantyfied via empirical fitting functions. A
forthcoming paper by Vazdekis et al.~(in preparation) will be devoted
to present and discuss the corresponding SSP model predictions (both
based on such fitting functions and from the SEDs in Vazdekis et
al.~2003) in comparison with galactic observational data.

Section~2 is focused on the definition of new line-strength indices
for the Mg\,{\sc i} line and the TiO bands and on the comparison with
those of previous work. The new index sensitivities to different
spectral resolutions, signal-to-noise ratios, flux calibration,
reddening uncertainties, and residuals of sky lines and telluric
absorptions are analyzed. Also, accurate formulae for estimating
random errors in the new index measurements are provided. After a
qualitative description of the Mg\,{\sc i} line and the TiO bands
strengths on the basis of the library stars, Section~3 is devoted to
the mathematical fitting procedure of the index strengths as functions
of the stellar atmospheric parameters, providing the significant
terms, coefficients, and statistics of the derived fitting
functions. A thorough analysis of the fitting function residuals and
possible error sources is herein presented, accounting for the
uncertainties in the input atmospheric parameters, the flux
calibration, and the effect of different [Mg/Fe] ratios in the
Mg\,{\sc i} line fitting functions. A qualitative comparison with MgI
and sTiO predictions based on theoretical work is presented in
Section~4. To conclude, Section~5 is reserved to discuss and summarize
the main contents and results of this paper.

\section{M\lowercase{g}I and T\lowercase{i}O spectroscopic indices}
\label{newindices}

\begin{figure*}
\centerline{\hbox{
\psfig{figure=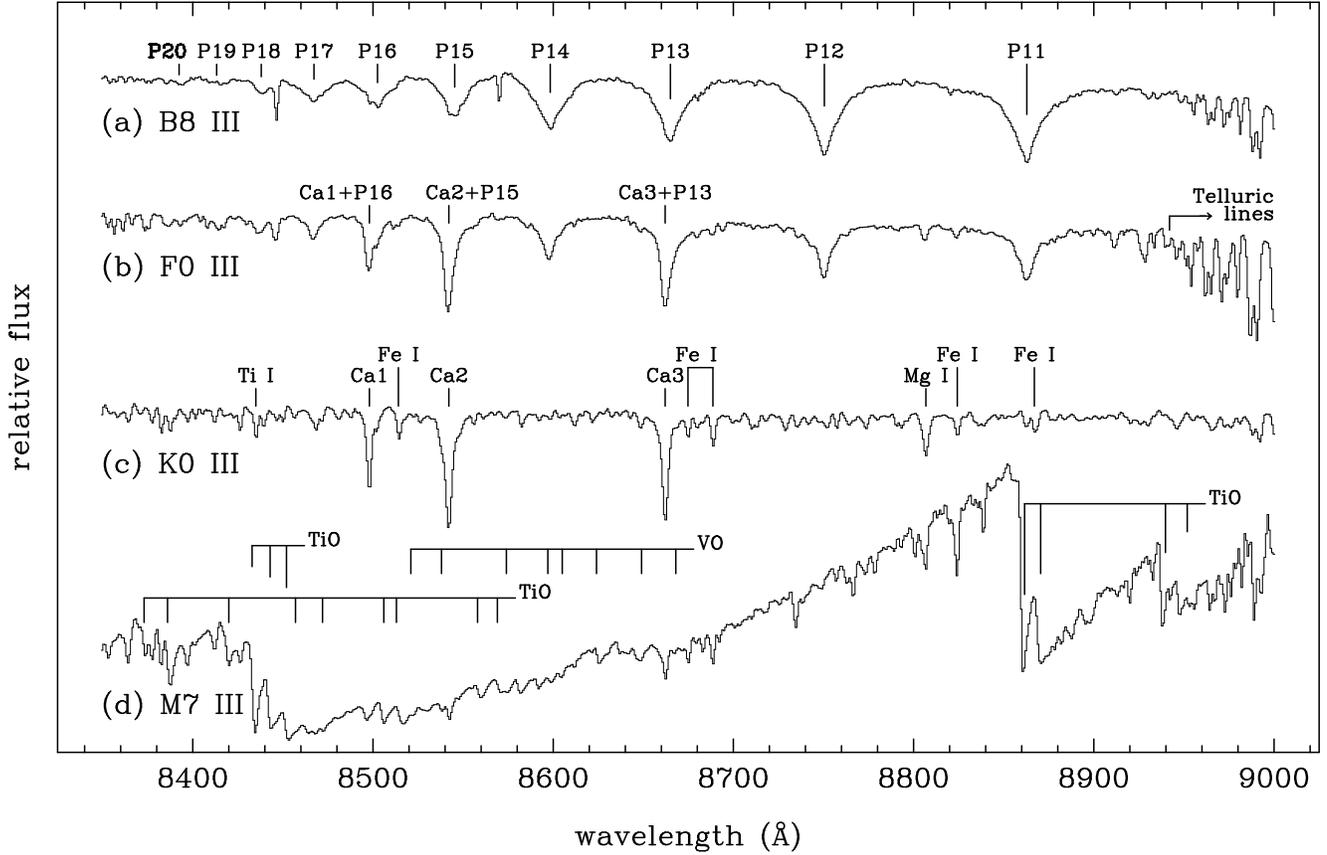}
}} 
\caption{Spectra of the stars HD 186568 (B8 III), HD 89025 (F0 III),
HD 216228 (K0 III) and HD 114961 (M7 III) in the spectral range of the
stellar library from CEN01a. The strongest features in this region are
marked: the Paschen Series (from P11 to P20), the Ca\,{\sc ii} triplet
(Ca1, Ca2 and Ca3), several metal lines and telluric absorptions. The
Mg\,{\sc i} line and the molecular bands of TiO and VO are indicated
in spectra (c) and (d) respectively.}
\label{identl}
\end{figure*}

Before focussing on the definition of new line-strength indices for
the measurement of the Mg\,{\sc i} and TiO spectral features, it is
worth making a brief description of the spectral range under
study. The existence of other absorption lines around the spectral
features of interest is indeed decisive for a proper location of the
index bandpasses. As in CEN01a, the strongest spectral features around
8600~\AA\ are labelled in Figure~\ref{identl} for a subsample of
representative spectral types; the MgI line and TiO bands concerning
this paper are indicated in spectra (c) and (d) respectively. It is
readily seen from that figure that the H Paschen series completely
dominates the spectra of the hottest stars. Its strength decreases
with the decrasing temperature at the time that several metal lines
become stronger. For G, K and early-M spectral types, the spectra are
mainly governed by the Ca\,{\sc ii} triplet, the Mg\,{\sc i} line and
many other Fe and Ti absorption lines. Finally, intermediate and
late-M types exhibit strong molecular bands of TiO and VO which
modulate the shape of the local continuum. See CEN01a and CEN02 for a
more detailed description of the above behaviours.

\subsection{Previous index definitions}
\label{previndex}

\begin{figure*}
\centerline{\hbox{
\psfig{figure=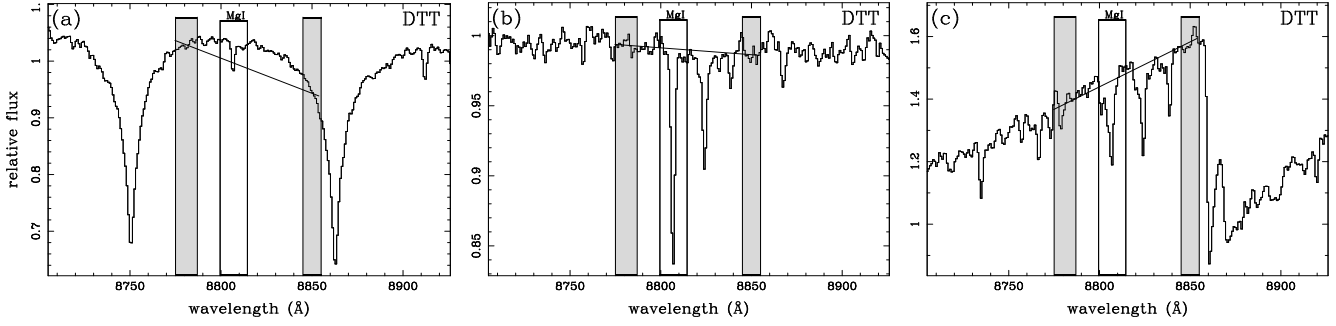,width=17.6cm}
}} 
\caption{\small Index definition by DTT for the Mg\,{\sc i} line
  over different spectral types. The spectra, taken from the stellar
  library of CEN01a, correspond to HD161817 (A2 VI; $a$), HD25329 (K1
  Vsb; $b$) and HD148783 (M6 III; $c$). Grey and open bands
  illustrate, respectively, the location of continuum and central
  bandpasses (see Table~\ref{prevMgIsTiO}). Solid lines represent the
  local pseudo-continua computed by means of error weighted
  least-squares fits to all the pixels in the continuum bandpasses.}
\label{mgiDTT}
\end{figure*}

\begin{figure*}
\centerline{\hbox{
\psfig{figure=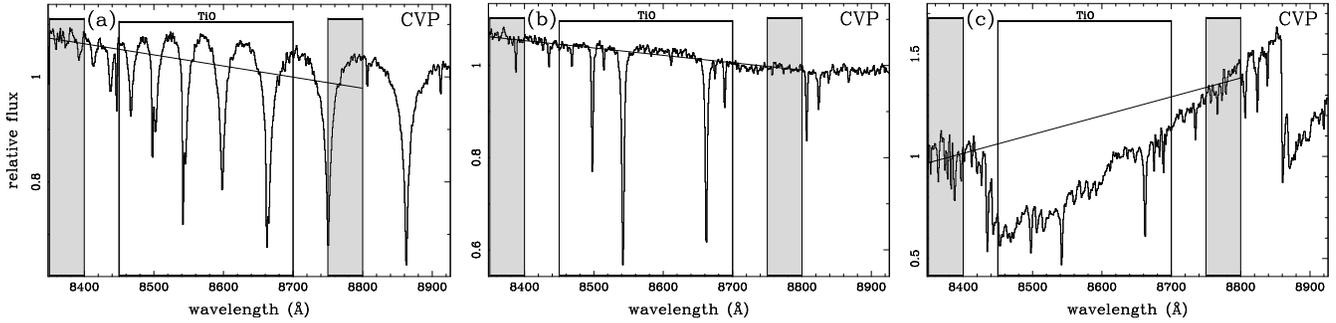,width=17.6cm}
}} 
\caption{\small Index definition for the TiO bands by CVP (TiO$_1$;
  see Table~\ref{prevMgIsTiO}) over different spectral types. Spectra,
  bandpasses color code, and solid lines are the same as in
  Fig.~\ref{mgiDTT}.}
\label{TiOCVP}
\end{figure*}

Despite most previous papers dealing with the present spectral range
were mainly focussed on the Ca\,{\sc ii} triplet, a few of them
already considered the Mg\,{\sc i} line at 8807~\AA\ and the TiO bands
at $\lambda\lambda$~8432, 8442, 8452~\AA\ and $\lambda\lambda$~8860,
8868~\AA. This was the case of D\'{\i}az, Terlevich \& Terlevich
(1989, hereafter DTT) and Carter, Visvanathan \& Pickles (1986,
hereafter CVP), who provided index definitions for the Mg\,{\sc i}
line and the TiO bands respectively. Table~\ref{prevMgIsTiO} lists the
bandpass limits of the corresponding indices as defined in the above
papers. In turn, Figures~\ref{mgiDTT} and~\ref{TiOCVP} illustrate the
suitability of these indices when measured over different spectral
types.

\begin{table}
\centering{
\caption{ Definition of previous and new indices for the near-IR
Mg\,{\sc i} line ($\lambda$~8807~\AA) and TiO bands
($\lambda\lambda$~8432, 8442, 8452~\AA; $\lambda\lambda$~8860,
8868~\AA). Codes for the references are as follow: CVP (Carter,
Visvanathan \& Pickles 1986), DTT (D\'{\i}az, Terlevich \& Terlevich
1989), TW (This work). Types $c$--$a$, $c$--$m$, $g$--$a$ and $g$--$s$
refer to classical--atomic, classical--molecular, generic--atomic and
generic--slope--like indices respectively.}
\label{prevMgIsTiO}
\vspace{4mm}
\begin{tabular}{@{}lccc@{}}
\hline\hline
Index & Type      & Central        & Continuum        \\ 
      &           & Bandpass (\AA) & Bandpasses (\AA) \\ 
\hline 
MgI (DTT)        &$c$--$a$     & 8799.5--8814.5 &  8775.0--8787.0 \\ 
                 &          &                &  8845.0--8855.0 \\ 
MgI (TW)         &$g$--$a$     & 8802.5--8811.0 &  8781.0--8789.0 \\ 
                 &          &                &  8831.0--8835.5 \\  \medskip           
                 &          &                &  8841.5--8846.0 \\
TiO$_{1}$ (CVP)  &$c$--$m$     & 8450.0--8700.0 &  8350.0--8400.0 \\
                 &          &                &  8750.0--8800.0 \\ 
TiO$_{2}$ (CVP)  &$c$--$m$     & 8890.0--9060.0 &  8790.0--8840.0 \\
                 &          &                &  9100.0--9150.0 \\ 
sTiO (TW)        &$g$--$s$     & none           &  8474.0--8484.0 \\ 
                 &          &                &  8563.0--8577.0 \\                          
                 &          &                &  8619.0--8642.0 \\                          
                 &          &                &  8700.0--8725.0 \\
                 &          &                &  8776.0--8792.0 \\      
\hline\hline
\end{tabular}
}
\end{table}

In the 90s, the indices defined by DTT were the most widely employed
to measure the strengths of the Ca\,{\sc ii} triplet\footnote{The
suitability of this and other Ca\,{\sc ii} indices is discussed in
CEN01a.} and the Mg\,{\sc i} lines. For the Mg\,{\sc i} spectral
feature they defined a {\it classical} atomic index that consists of a
central bandpass enclosing the Mg\,{\sc i} line, and two continuum
bandpasses located at both sides ---blue and red--- of the central
one. In spite of being a well-defined index for F$-$K spectral types,
it suffers from two main limitations: i) the presence of strong
Paschen lines (P11 and P12) in early spectral types makes the derived
pseudo-continuum to be unreliable (Figure~\ref{mgiDTT}a), and ii) the
proximity of the red continuum bandpass to the TiO break around
8600~\AA\ makes the index to be very sensitive to spectral resolution
and velocity dispersion broadening (Figure~\ref{mgiDTT}c). This is
particularly critical for the integrated spectra of galaxies in which
TiO bands may be prominent (see e.g.~the SEDs predicted in VAZ03), and
typical velocity dispersions are above $\sim 100~{\rm km s}^{-1}$. In
any case, it is fair noting the intrinsic difficulty of defining a
Mg\,{\sc i} line-strength index in a spectral region dominated by
strong absorption lines. This is particularly evident in the case of
the earliest spectral types (see Figure~\ref{mgiDTT}a), for which the
Mg\,{\sc i} line is wery weak and the wings of the Paschen lines are
blended thus decreasing the true continuum level.

The work by CVP defined {\it classical} molecular indices to measure
the strength of the TiO bands. Again, the index consists of two
continuum bandpasses and a very wide central bandpass for the TiO
break at $\sim 8450$~\AA\ (see Table~\ref{prevMgIsTiO}). Because of
the width of the central bandpass, the index turns out to be sensitive
the Ca\,{\sc ii} triplet and the Paschen lines for those spectral
types in which the above features dominate the spectrum
(Figures~\ref{TiOCVP}a,b). Such a Ca and H contamination for most
spectral types is obviously not desired when trying to understand the
behaviour of the TiO bands with the stellar atmospheric parameters, as
it would translate into a blurring of age and metallicity effects if
the TiO index were used as a tool for stellar population diagnostics.

\subsection{New index definitions}
\label{newindexdef}

In CEN01a we introduced new type of line-strength indices, namely {\it
generic} indices, which allow the definition of an arbitrary but
precise number of continuum bandpasses to derive the pseudo-continuum
level. It is computed as an error-weighted, least-squares linear fit
to all the pixels of these continuum bandpasses. Generic indices also
allow the inclusion of various bandpasses for adjacent spectral
features, which are thus measured simultaneously -using different
relative weights- under the same pseudo-continuum. As we report in
that paper, the above improvements are highly advantageous in regions
densely populated by other spectral features, telluric lines or strong
sky emission lines, as it is the case for the near-IR spectral
range. CaT, PaT and CaT$^*$ are examples of generic indices for the
Ca\,{\sc ii} triplet and three lines of the H Paschen series (see
details in CEN01a).

The new generic indices that we characterize here, MgI and sTiO, were
formerly measured in Cenarro et al.~(2003) for a sample of early-type
galaxies. We devote the current paper to provide full details on their
definition, sensitivities, and behaviour with the stellar atmospheric
parameters.

\subsubsection{The MgI index}

\begin{figure*}
\centerline{\hbox{
\psfig{figure=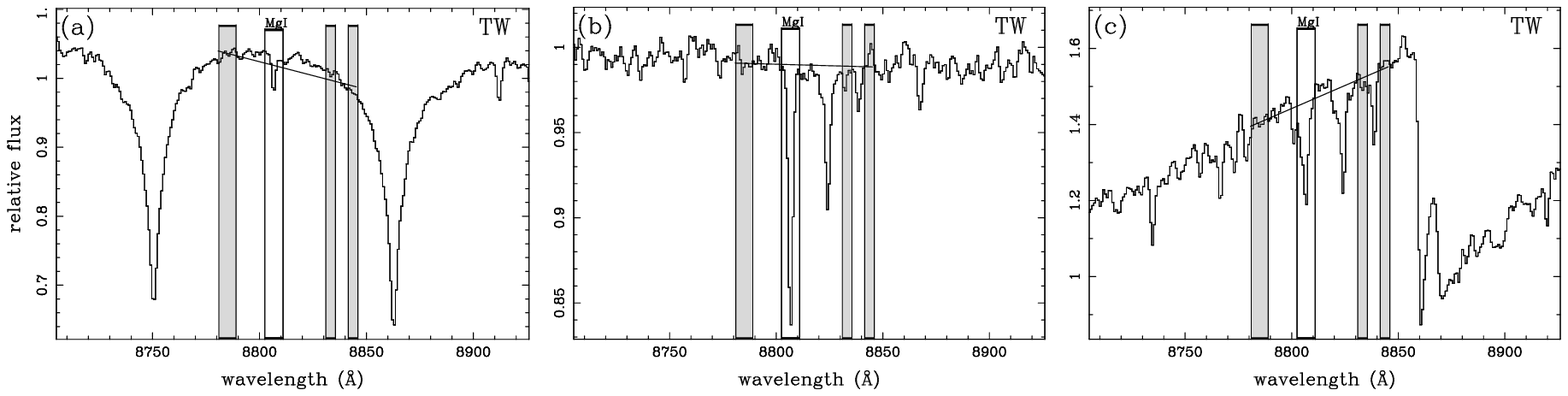,width=17.6cm}
}} 
\caption{\small New index definition (this work; TW) for the Mg\,{\sc
    i} line, MgI, over the same spectral types of
  Fig.~\ref{mgiDTT}. Again, grey and open bandpasses illustrate,
  respectively, the location of continuum and central bandpasses (see
  Table~\ref{prevMgIsTiO}), whilst solid lines represent the local
  pseudo-continua derived from error weighted least-squares fits to
  all the pixels in the continuum bandpasses.}
\label{mgiCEN}
\end{figure*}

The MgI index has been defined as a generic index consisting of three
continuum bandpasses and one spectral-feature bandpass for the
Mg\,{\sc i} line. Figure~\ref{mgiCEN} illustrates the Mg\,{\sc i}
index defined in this work when measured over the same spectra as in
Fig.~\ref{mgiDTT}, and the bandpasses limits are listed in
Table~\ref{prevMgIsTiO}. The location and width of these bandpasses
were established in order to derive a reliable pseudo-continuum for
all the spectral types even when the spectra are broadened up to
300~${\rm km s}^{-1}$. Also, since we are interested in measuring MgI
on broadened galaxy spectra, we ensured that the TiO break at
$\lambda$~8860\,\AA\ was not affecting the pseudo-continuum
level. Note however that, because of the problem reported in the
previous section, the pseudo-continuum derived for early spectral
types is still slightly below the true level (Fig.~\ref{mgiCEN}a).
With the aim of defining a reliable indicator of Mg abundance, we
tried to avoid as much as possible the presence of other metal lines
within the spectral-feature band (e.g. see Fig.~\ref{mgiCEN}c). This
is why we preferred to define a quite narrow characteristic bandpass,
even though it increases the sensitivity of the index to the spectral
resolution. A reasonable compromise between both requirements was
finally established.

\subsubsection{The sTiO index}
\label{indexstiodef}

For the measurement of the TiO bands we introduce a new type of
generic index which will be referred to as {\it slope} index. Slope
indices are defined as the ratio between the pseudo-continuum values
at the central wavelengths of any two continuum bandpasses. In this
sense, they can be considered as a measurement of the local
pseudo-continuum slope. It is important to note that, although slope
indices just consider the pseudo-continuum values at the center of two
continuum bandpasses, the total number of continuum bandpasses driving
the pseudo-continuum level can be as large as desired. In practice one
actually performs an error-weighted least-squares linear fit to all
the pixels within the full set of considered continuum bandpasses.
After that, the linear fit is evaluated at the central wavelengths of
the first and last of those bandpasses.

This new concept of index was conceived with the aim of measuring the
slope of the continuum around the Ca\,{\sc ii} triplet, mainly
governed by molecular absorptions (TiO and VO) in mid and late-M
spectral types. Given that the location of the five continuum
bandpasses for the indices CaT, PaT or CaT$^*$ leads to a reliable
pseudo-continuum for all the spectral types, we took advantage of the
previous indices to define the slope index sTiO. In particular, it is
defined as the ratio between the pseudo-continuum values,
$C(\lambda)$, at the central wavelengths of the reddest and bluest
continuum bandpasses (see Table~\ref{prevMgIsTiO}), that is,

\begin{equation}
{\rm sTiO} = \frac{C(\lambda8784.0)}{C(\lambda8479.0)} .
\label{sTiOdef}
\end{equation}

At variance with other spectrophotometric indices, sTiO is potentially
sensitive to flux calibration uncertainties (hence extinction effects)
given that its continuum bandpasses spread over more than 300\,\AA. On
the other hand, it is particularly unsensitive to low signal-to-noise
ratios as the slope gets robustly constrained by 5 bandpasses that
overall cover 88\,\AA. These and other effects will be discussed in
Section~\ref{sensitivity}.

\subsubsection{MgI and sTiO measurements for the library stars}
\label{newindexmeasurements}

The new indices have been measured for all the spectra in CEN01a at
the nominal resolution of the library, that is, FWHM$ = 1.5$\,\AA\ or
$\sigma = 22.2$\,km~s$^{-1}$. Table~\ref{macrotab} lists the index
measurements and their errors, which account for the photon noise and
radial velocity uncertainties, the later including typical errors in
wavelength calibration. This database is also available at \\ {\tt
http://www.ucm.es/info/Astrof/ellipt/MgIsTiO.html}\\

The actual measurements have been performed with {\tt indexf} (Cardiel
2007), a C++ program specially written to compute atomic, molecular,
break, generic-atomic, generic-break and slope indices in
wavelength-calibrated FITS spectra. This program is available at\\
{\tt http://www.ucm.es/info/Astrof/software/indexf}

\begin{figure*}
\centerline{\hbox{
\psfig{figure=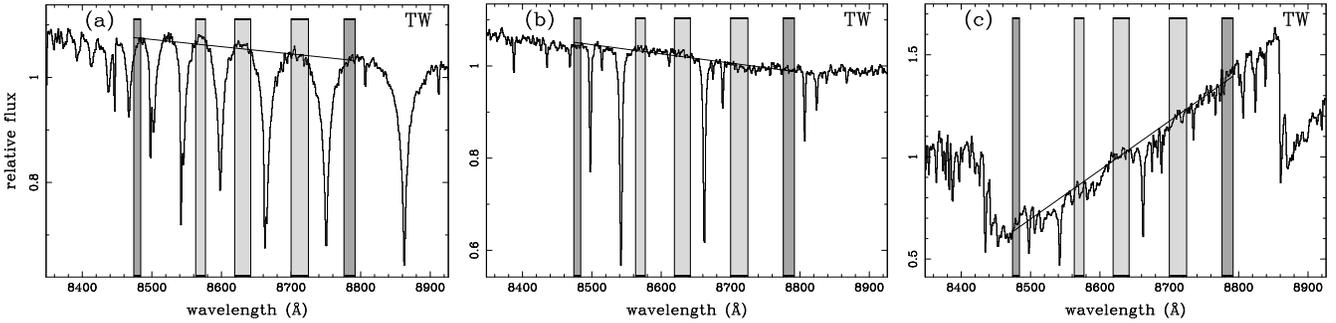,width=17.6cm}
}} 
\caption{\small New index definition (this work; TW) for the TiO
  bands, sTiO, over the same spectral types of Fig.~\ref{mgiDTT}. The
  five continuum bandpasses are those corresponding to the CaT indices
  in CEN01a. The two bandpasses that define the sTiO value (see
  Eq.~\ref{sTiOdef}) are illustrated in dark grey.}
\label{sTiOCEN}
\end{figure*}

\subsection{Conversions between the new and previous index systems}
\label{calibrations}

\begin{table}
\centering{
\caption{Calibrations between different index systems.  $\sigma_{\rm
r.m.s}$: unbiased standard deviation of the fit. N: Number of stars in
the fit. {\it T}$_{{\rm eff}}$ (K): Effective temperature region where
the calibration was obtained. These calibrations are graphically
displayed in Fig.~\ref{compsplot}.}
\label{compstab}
\begin{tabular}{@{}lccc@{}}
\hline
\multicolumn{1}{c}{Calibrations}& $\sigma_{\rm r.m.s}$ & $N$ & {\it T}$_{{\rm eff}}$ (K)\\
\hline
MgI = 0.119 + 0.864 MgI(DTT) & 0.07 & 560 & 2750--6300 \\ \medskip
sTiO = 0.822 + 3.397 TiO$_{1}$(CVP) & 0.07 & 29 & 2750--3700 \\
MgI(DTT) = MgI(DTT)$_{{\rm DTT}}$ & 0.16 & 100 & 3425--6800 \\
\hline
\end{tabular}
}
\end{table}

\begin{figure*}
\centerline{\hbox{
\psfig{figure=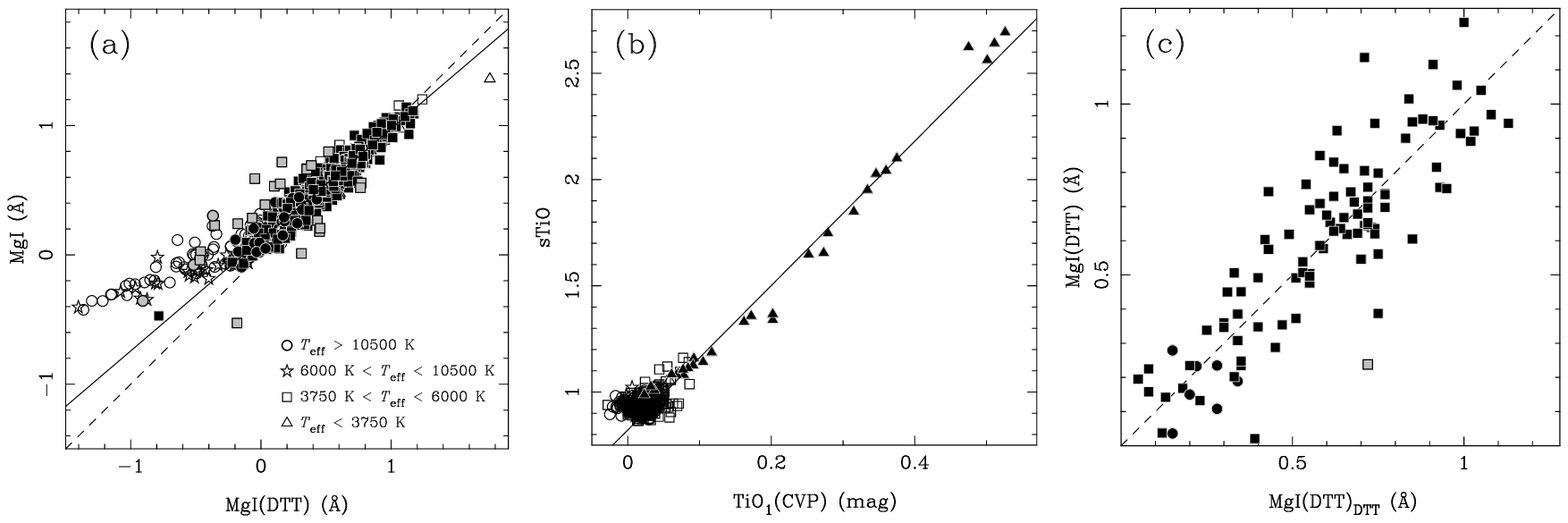}
}}
\caption{Comparison between the different systems of indices. Diagrams
(a) and (b) compare the new indices MgI and sTiO with the
corresponding MgI (by DTT) and TiO$_{1}$ (by CVP), both measured in
this work over the 706 stars of our stellar library (CEN01a). Diagram
(c) shows the MgI by DTT measured in our and their spectra (MgI(DTT)
and MgI(DTT)$_{{\rm DTT}}$ respectively) for the subsample of stars in
common. Symbol types, indicating different ranges of effective
temperature, are given in panel (a). The dashed line in panels (a) and
(c) shows the one-to-one relation. Grey symbols are stars deviating
more than $3\sigma$ from the fitted relation, whereas open symbols
refer to those stars with effective temperatures outside the fitted
range. The solid line marks the most significant fit to the symbols in
black (see Table~\ref{compstab}).}
\label{compsplot}
\end{figure*}

For those readers interested in transforming old system data of
indices into the new ones (or vice-versa), this section provides a set
of calibrations to make conversions from one system to the 
other. Table~\ref{compstab} lists the derived relations and
Figure~\ref{compsplot} illustrates the corresponding fits.

Using the 706 library stars from CEN01a, we have compared the
measurements of the indices MgI(DTT) and TiO$_{1}$(CVP) with the ones
corresponding to the new MgI and sTiO. The calibrations were computed
by means of error weighted least-squares fits to a straight line that,
in both cases, turned out to be statistically significant. The
location and width of the MgI(DTT) bandpasses (see
Section~\ref{previndex}) causes hot stars to depart from the general
trend of the rest of the library stars. In order to ensure the quality
of the fit, we have restricted the range of the calibration excluding
from the fit those stars with high temperatures
(Fig.~\ref{compsplot}a). The calibration between the TiO indices
(Fig.~\ref{compsplot}b) was derived just considering those stars cold
enough to exhibit molecular bands in their spectra. Otherwise the fit
would be strongly dominated by the behaviour of earlier spectral types
for which the indices do not measure TiO. In all fits, stars within
the valid range of $T_{\rm eff}$ but deviating more than $3\sigma$
from the fitted relation were also rejected.

Finally, for a subsample of stars in common with the stellar library
from DTT, we have compared the MgI(DTT) values given in DTT
(MgI(DTT)$_{{\rm DTT}}$) with the ones measured over our spectra. Note
that, since the measured index is the same in both cases, systematic
differences between the two sets of measurements could only arise from
differences in their spectrophotometric systems.  However, no
significant differences have been found (Fig.~\ref{compsplot}c).

To conclude, it is important to remind the reader that the two first
calibrations in Table~\ref{compstab} should only be applied once the
spectra are on the same spectrophotometric system (that is, equally
flux calibrated and at the same spectral resolution) as the stellar
library of CEN01a. If that is not the case, a prior calibration like
the last one in Table~\ref{compstab} should be applied before.

\subsection{Sensitivities of the new indices to different effects}
\label{sensitivity}

This section is devoted to characterize the sensitivity of the indices
MgI and sTiO to the signal-to-noise ratio, the velocity dispersion
broadening (or spectral resolution), relative flux calibration,
extinction effects, and sky subtraction residuals.

\subsubsection{Signal-to-noise ratio}
\label{snsens}

In CEN01a, it was presented a thorough study on the computation of
random errors ---arising from photon noise--- for generic indices. We
refer the reader to the Appendix~A2 of that paper for full details
about the procedure (see also Cardiel et al.~1998). In those papers,
it was demonstrated that it is possible to estimate the predicted
random errors of atomic indices as a function of the signal-to-noise
(S/N) ratio per angstrom ---hereafter $S\!N(\rm\AA)$---, by means of
analytical expressions of the form
\begin{equation}
\sigma[{\cal I}_{\rm a}] \simeq
  \displaystyle \frac{c_1-c_2\;{\cal I}_{\rm a}}{S\!N(\rm\AA)},
\label{errorvssn}
\end{equation}
where [${\cal I}_{\rm a}$] refers to any atomic index (either
classical or generic), and the coefficients $c_1$ and $c_2$ ---which
depend on the index definition--- can be computed analytically for
classical indices, and empirically for generic ones (see Appendix~3 in
CEN01a).

\begin{figure}
\centerline{\hbox{
\psfig{figure=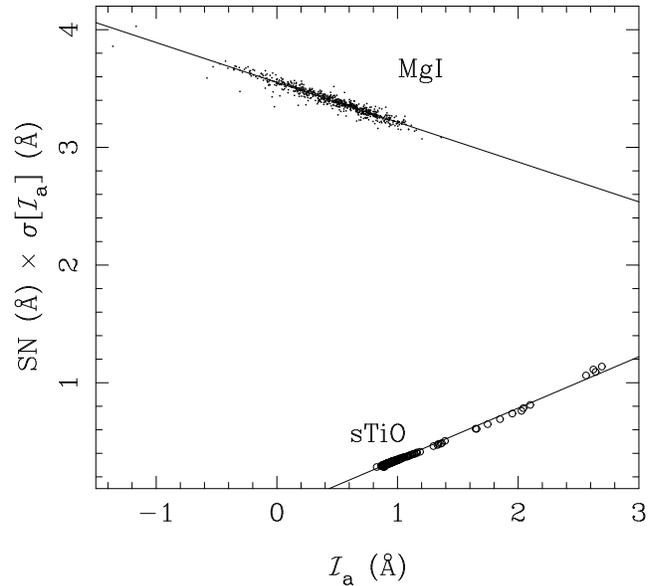,width=8.4cm}
}}
\caption{\small Empirical estimation of the constant factors $c_1$ and
$c_2$ in Eqs.~(\ref{rel_appr_mgi}) and (\ref{rel_appr_stio}), for MgI
and sTiO, respectively. The least-squares fits to straight lines have
been performed rejecting data iteratively outside the 99.73 per cent
confidence level.}
\label{mgi-stio-SN}
\end{figure}

Following the empirical approach for our generic indices,
Figure~\ref{mgi-stio-SN} displays the product $S\!N(\rm\AA) \times
\sigma[{\cal I}_{\rm a}]$ versus the index value for all the library
stars in CEN01a. As expected for generic atomic indices, MgI exhibits
a clear linear relationship, thus supporting previous results that
Eq.~\ref{errorvssn} is indeed a good approach to the S/N dependence of
random errors. Interestingly, although sTiO is not an atomic index, it
is also found to follow a nice linear behaviour when included in
Fig.~\ref{mgi-stio-SN}, despite it exhibits an opposite trend (what is
understandable as slope indices are conceptually different to atomic
indices; see Section~\ref{indexstiodef}. Note also that sTiO has no
units, so the labels of the axes are not strictly correct in this
case). A least-squares fit to all data provides
\begin{equation}
\begin{array}{@{}r@{\;}c@{\;}l}
\sigma[\rm MgI] & \simeq &
 \displaystyle\frac{3.552-0.3384\;\rm MgI}{S\!N(\rm\AA)}, \\
\end{array}
\label{rel_appr_mgi}
\end{equation}
and
\begin{equation}
\begin{array}{@{}r@{\;}c@{\;}l}
\sigma[\rm sTiO] & \simeq &
 \displaystyle\frac{-0.08914+0.4366\;\rm sTiO}{S\!N(\rm\AA)}. \\
\end{array}
\label{rel_appr_stio}
\end{equation}
As expected, for MgI, $c_1$ is one order of magnitude larger than
$c_2$, thus reinforcing the idea that photon noise errors in generic
atomic indices are barely dependent on the index values. However, this
is not the case for slope indices like sTiO. In fact, $c_2$ dominates
the error dependence, in the sense that the larger the index, the
larger the error.

Based on the above equations we estimate that, for the typical indices
of an old, solar-metallicity SSP (VAZ03), it is required $S\!N(\rm\AA)
\sim 50$\,\AA$^{-1}$ and $\sim 3$\,\AA$^{-1}$ to measure,
respectively, MgI and sTiO with a 10 per cent uncertainty. It is
therefore clear that, while sTiO is optimized to be measured on low
signal-to-noise ratio spectra (a potential application for high
redshift galaxies and extragalactic globular clusters is immediately
derived), MgI demands relatively high quality spectra to drive
reliable results. This is unavoidably the price one has to pay for
defining MgI as a ``pure'' Mg indicator, with a very narrow central
bandpass that prevents from the contamination with other nearby lines.

\subsubsection{Spectral resolution}
\label{subsec:veldisp}

\begin{figure}
\psfig{figure=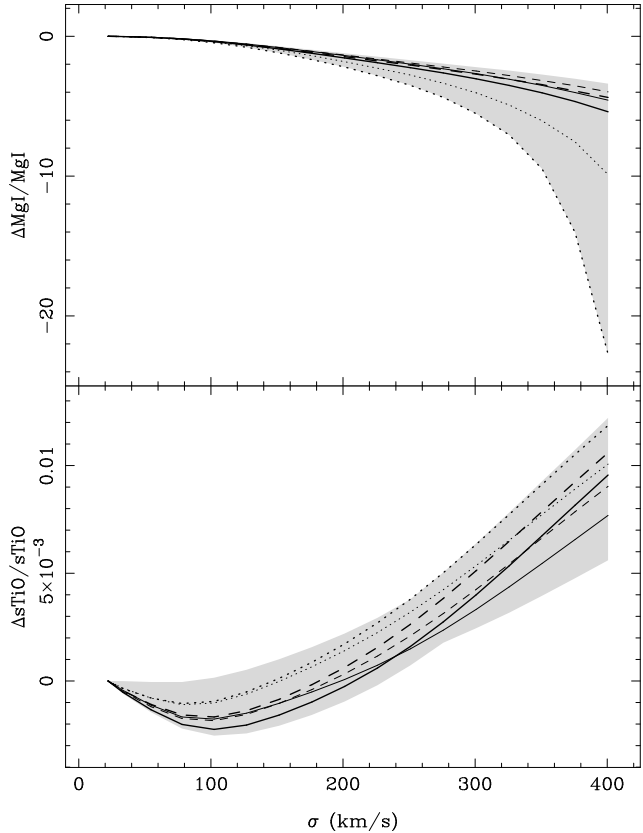,width=8.4cm}
\caption{Broadening correction $\Delta$MgI/MgI and $\Delta$sTiO/sTiO
as a function of total $\sigma$. Corrections have been computed by
convolving model spectra by VAZ03 with Gaussians from $\sigma=25$
km~s$^{-1}$ up to $\sigma=400$ km~s$^{-1}$ in steps of 25
km~s$^{-1}$. $\Delta I/I$ is zero for $\sigma = $22.2 km s$^{-1}$ (the
spectral resolution of the stellar library in CEN01a). Different line
types are employed to illustrate the broadening corrections for a set
of representative models of different ages, metallicities, and IMFs:
solid (12.6\,Gyr, [M/H]$=0.0$), dotted (1.0\,Gyr, [M/H]$=0.0$), and
dashed (12.6\,Gyr, [M/H]$=-0.7$), with thick and thin lines
corresponding to IMF slopes of $\mu =1.3$ (Salpeter) and 2.8
respectively. The polynomial coefficients of such broadening
corrections are given in Table~\ref{polcoefdisp}.  The grey area
illustrates the region covered by the broadening corrections obtained
for the whole set of SSP spectra of VAZ03.}
\label{fig:pols_sigma}
\end{figure}

With the aim of studying the sensitivity of MgI and sTiO to the
spectral resolution or galaxy velocity dispersion broadening
($\sigma$), we broadened the whole set of SSP model spectra of VAZ03
by convolving with Gaussians of $\sigma$ varying from 25 up to
$400$\,km~s$^{-1}$, in steps of 25 km~s$^{-1}$. The indices were thus
measured for the full set of broadened spectra and, for each model, we
fitted a third-order polynomial to the relative changes of the index
values as a function of velocity dispersion,

\begin{equation} 
{\frac{{\cal I}(\sigma) - {\cal I} (\sigma_{0})}{{\cal I}(\sigma)} = a
+ b\sigma + c\sigma^{2} + d\sigma^{3} \equiv p(\sigma)},
\label{poly}
\end{equation}

\noindent 
where $\sigma_{0}=22.2$ km~s$^{-1}$ is the nominal resolution of the
VAZ03 models (FWHM = 1.50 \AA), and $\sigma$ is any generic resolution
in the same units. Note that Eq.~\ref{poly} is computed as a function
of the overall spectral resolution, so that for $\sigma =
\sigma_{0}$, $p(\sigma_{0}) = 0$.

Table~\ref{polcoefdisp} provides the derived coefficients for MgI and
sTiO measured over a number of representative SSP
models. Figure~\ref{fig:pols_sigma} illustrates the obtained
$\Delta$MgI/MgI and $\Delta$sTiO/sTiO values for this set of
representative models. The grey region represents the locus of
broadening corrections for the whole SSP model spectral library.

It is worth noting that, thanks to the high stability of the continuum
bandpasses, the index sTiO is formally insensitive to broadening, with
largest corrections being only $\sim 1$ per cent at $\sigma \sim
400$\,km~s$^{-1}$. On the other hand, MgI turns out to be quite
dependent on spectral resolution. This is due to the small width of
the central bandpass ---chosen this way to preserve as much as
possible the Mg abundance sensitivity--- and to the intrinsic weakness
of the Mg\,{\sc i} line (EW $< 1$\,\AA). In fact, at $\sigma =
300$\,km~s$^{-1}$, Mg\,{\sc i} decreases by $\sim 78.2 \pm 6.6$ per
cent w.r.t. the value at $\sigma_{0} = 22.2$ km~s$^{-1}$, with
uncertainties accounting for the different broadenings derived from
the full set of SSPs. This effect decreases down to $\sim 56.8 \pm
0.6$ per cent at $\sigma = 200$\,km~s$^{-1}$, which can be considered
as a reasonable limiting resolution to neglect broadening correction
differences due to distinct SSP templates.

As it was already discussed in VAZ03 for the Ca\,{\sc ii} triplet
indices, the use of SSP model spectra to match the effects of galaxy
broadening is a better approach than the use of single stellar
spectra, as the later shows even large variations among different
spectral types. In any case, the user should still keep in mind that
systematic differences ---arising, for instance, from different
abundance ratios or local flux calibration mismatches--- may exist
between real galaxies and the best matched SSP models. For this
reason, whenever it is feasible, an alternative approach is that of
broadening the galaxy spectra and the SSP models up to the largest
spectral resolution of the galaxy sample. The success of this approach
for the MgI and sTiO indices of elliptical galaxies over a range in
mass was illustrated in Cenarro et al.~(2003).

\begin{table*}                                                              
\centering{
\caption{Coefficients of the broadening correction polynomials $\Delta
  {\cal I}/{\cal I} = a + b\sigma + c\sigma^{2} + d\sigma^{3}$ for MgI
  and sTiO, for a representative set of Vazdekis et al.~(2003) SSP
  models of different IMF slopes $\mu$, metallicities [M/H], and
  ages. $\Delta {\cal I}/{\cal I}$ is zero for $\sigma = $22.2 km
  s$^{-1}$, the spectral resolution of the stellar library in CEN01a.}
\label{polcoefdisp}                                                            
\begin{tabular}{@{}crrrr@{}lr@{}lr@{}lr@{}l@{}}
\hline
 ${\cal I}$ & $\mu$ & [M/H] & Age (Gyr) &
\multicolumn{2}{c}{$a(\times10^{-3})$} & 
\multicolumn{2}{c}{$b(\times10^{-5})$} & 
\multicolumn{2}{c}{$c(\times10^{-7})$} & 
\multicolumn{2}{c}{$d(\times10^{-9})$}\\
\hline     
sTiO&1.3&\    0.0&\   1.0&      0.&909   &   $-4.$&782  &     3.&179  &   $-0.$&334\\
    &1.3&\    0.0&\  12.6&      1.&421   &   $-7.$&219  &     3.&764  &   $-0.$&368\\
    &1.3&\ $-0.$7&\  12.6&      0.&895   &   $-4.$&648  &     2.&828  &   $-0.$&262\\
    &2.8&\    0.0&\   1.0&      0.&753   &   $-3.$&941  &     2.&536  &   $-0.$&241\\
    &2.8&\    0.0&\  12.6&      1.&064   &   $-5.$&434  &     2.&947  &   $-0.$&296\\ 
    &2.8&\ $-0.$7&\  12.6&      1.&145   &   $-5.$&879  &     3.&333  &   $-0.$&341\\
&&&&&&&&&&&\\
MgI &1.3&\    0.0&\   1.0&   1030.&686   &$-5544.$&545  &  4297.&828  &$-1061.$&284\\
    &1.3&\    0.0&\  12.6&     61.&862   & $-207.$&795  &$-259.$&635  &   $-4.$&898\\
    &1.3&\ $-0$.7&\  12.6&     44.&553   & $-124.$&534  &$-349.$&063  &   $27.$&095\\
    &2.8&\    0.0&\   1.0&    225.&317   &$-1124.$&182  &   539.&861  & $-215.$&236\\
    &2.8&\    0.0&\  12.6&     42.&790   & $-127.$&770  &$-295.$&117  &   $10.$&880\\
    &2.8&\ $-0$.7&\  12.6&     37.&606   &  $-95.$&207  &$-340.$&763  &   $29.$&619\\
\hline
\end{tabular}                                                              
}                                                                          
\end{table*}

\subsubsection{Flux calibration and reddening correction}
\label{fluxcalx}

There is not a simple recipe to quantify analytically the sensitivity
of an index to the uncertainties in the relative flux calibration, as
it does not only depend on the index definition itself, but also on
the uncertainties of the response curves derived for a given observing
run. In general, the sensitivity of any spectroscopic index to flux
calibration gets larger as the spectral coverage of its sidebands
increases. In this sense, it is clear that sTiO is particularly
sensitive to systematics in the flux calibration and the reddening
correction, as it basically measures the slope of the pseudo-continuum
in a spectral window of $\sim 300$\,\AA. For instance, assuming the
Galactic interstellar extinction of Fitzpatrick (1999), the effect of
reddening in the sTiO index is given by
\begin{equation}
\displaystyle \frac{{\rm sTiO}_{\rm ext}-{\rm sTiO}_0}{{\rm sTiO}_0} =
0.087\times E(B-V),
\label{eqextinction}
\end{equation}
where sTiO$_{\rm ext}$ is the reddened index value (for a colour
excess of $E(B-V)$ and R = 3.1) and sTiO$_0$ the extinction corrected
one. This means that to ensure systematics in the sTiO index below
$1\%$, reddening correction uncertainties should not be larger than
$E(B-V) \sim 0.1$\,mag. MgI, however, is basically stable under flux
calibration variations and extinction corrections.

As it will be discussed in Section~\ref{uncfluxcal} with the aim of
explaining the observed fitting function residuals, it is important to
note that the main source of random errors for the sTiO index turns
out to be the random uncertainties in the determination of the
response curve. This stresses the importance of an accurate flux
calibration and extinction correction before performing any meaningful
comparison between evolutionary synthesis model predictions and
measured spectra.

\subsubsection{Sky emission lines and telluric absorptions}
\label{subsky}

Sky emission lines ---produced by the OH radical (e.g.~Rousselot et
al.~2000)--- and telluric absorptions ---due to water vapour and other
molecules (e.g. Stevenson 1994, Chmielewski 2000)--- are common
features at the near-IR spectral range. A careful data reduction is
necessary for a proper removal of these effects, although a detailed
description of specific techniques is out of the scope of this
paper. Rather than that, we just aim to compare qualitatively the
potential sensitivities of our indices to the above contaminations. An
absolute study is not possible at this point, as it would strongly
depend on the quality of the final spectra.

Because of the narrow index definition and the intrinsic weakness of
the Mg\,{\sc i} line, MgI is quite more sensitive to sky line and
telluric absorptions residuals than sTiO. In principle, since the sTiO
slope is well constrained by five continuum bandpasses, a few
deviating pixels in its continuum bandpasses would not alter the index
value significantly (actually, generic indices are particularly
insensitive to sky line residuals; Section~\ref{newindexdef}). A
different situation would be that in which a strong ---not properly
removed--- telluric absorption is affecting the red ---or blue---
sides of the index definition. In this case, the local slope of the
continuum would be fictitiously biased and the index would be highly
unreliable. Fortunately, this is not the case for spectra at the local
rest-frame (like the stellar library of CEN01a). In any case, it is
worth noting that the strong telluric absorption at $\lambda \ga
8940$\,\AA\ may affect the MgI and sTiO indices of objects with radial
velocities larger than $\sim 3200$ and $5000\,{\rm km~s}^{-1}$
respectively.

\section{The dependence of the M\lowercase{g}I line and T\lowercase{i}O bands on the stellar atmospheric parameters}
\label{stellib}

\subsection{Qualitative behaviour}
\label{qualitative}

\begin{figure}
\centerline{\hbox{
\psfig{figure=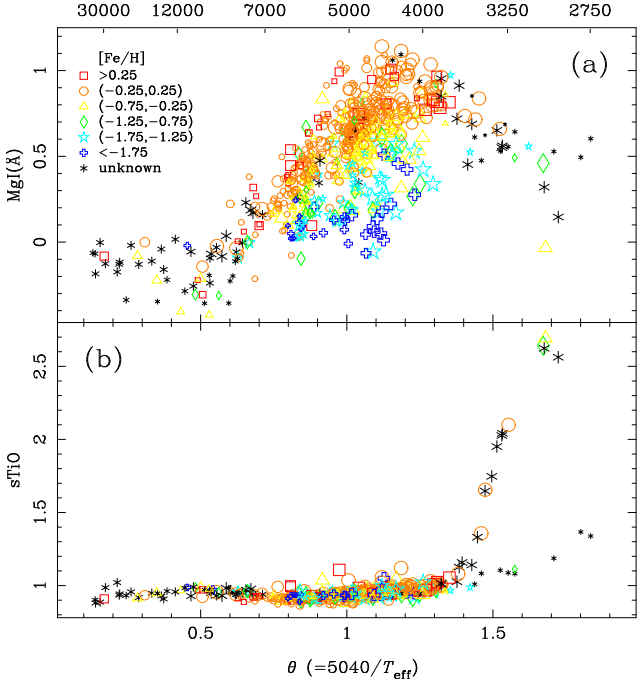,width=8.4cm}
}} 
\caption{\small MgI and sTiO indices versus $\theta (\equiv
  5040/T_{\rm eff})$ for the whole stellar library in
  CEN01a. Different symbols are used to indicate different ranges of
  metallicities (as in the key), while sizes are related with surface
  gravity, in the sense that the smaller the symbol (dwarfs), the
  higher the gravity. On the top, the effective temperature scale is
  given.}
\label{MgIsTiOtheta}
\end{figure}

As a first step to understand the behaviour of the Mg\,{\sc i} line
and TiO bands as a function of the stellar atmospheric parameters,
this section describes, from a qualitative point of view, the effects
of effective temperature ($T_{\rm eff}$ or $\theta \equiv 5040/T_{\rm
eff}$), surface gravity ($\log g$) and metallicity ([Fe/H]) on the
strength of both spectral features. Figure~\ref{MgIsTiOtheta}
illustrates the MgI and sTiO indices versus $\theta$ for all the stars
in CEN01a.

\subsubsection{The stellar atmospheric parameters}
\label{stelparam}

The stellar atmospheric parameters employed throughout this work are
those derived in CEN01b, where, after an exhaustive compilation from
several hundreds of bibliographic sources, the different sources were
calibrated and corrected onto the system established by Soubiran, Katz
\& Cayrel (1998) to end up with a homogeneous set of atmospheric
parameters (see details in CEN01b). Although more recent
determinations of $T_{\rm eff}$, $\log g$, and/or [Fe/H] have appeared
since 2001 for some library stars, we preferred to keep the parameters
quoted in CEN01b to preserve full consistency with the CaT, PaT, and
CaT$^{*}$ fitting functions given in CEN02, as well as with the
corresponding SSP SEDs predicted in VAZ03.  It is worth noting that,
in Cenarro et al.~(2007), an updated extension of the compiling work
in CEN01b was carried out for MILES, which in turn includes 403 stars
in common with CEN01a. Using all those stars in CEN01a for which any
of the three atmospheric parameters in Cenarro et al.~(2007) was
updated with respect to CEN01b (194 stars in $T_{\rm eff}$, 133 in
$\log g$, and 116 in [Fe/H]), we determined that the differences
between the old and new final determinations are null on average, with
r.m.s.~standard deviations of $\sigma T_{\rm eff} \sim 72$\,K, $\sigma
\log g \sim 0.22$\,dex, and $\sigma$[Fe/H]$ \sim 0.1$\,dex. Since
there exist not statistically significant offsets between both
datasets in any of the three atmospheric parameters that may lead to
systematic effects, and the scatter of the distributions is smaller
than the typical uncertainties in the atmospheric parameters (see
Section~\ref{uncatmpar}), we are confident that using the parameters
in CEN01b is not compromising at all the quality of the results that
we present in this work.

\subsubsection{The Mg\,{\sc i} line}
\label{qualmgi}

\begin{figure}
\centerline{\hbox{
\psfig{figure=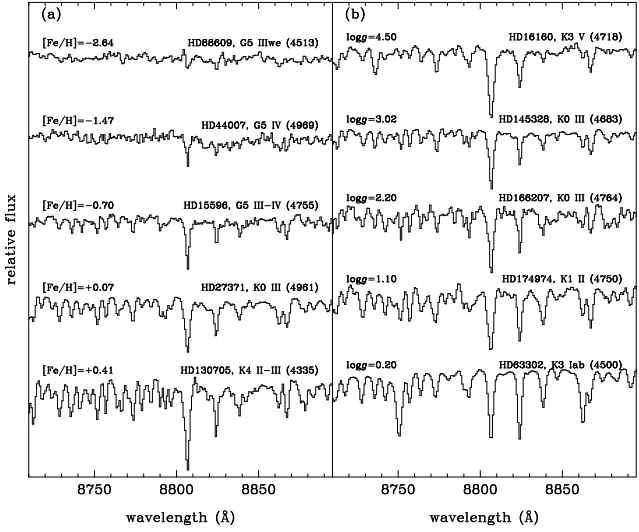,width=8.4cm,angle=-90}
}} 
\caption{\small Metallicity and gravity effects on the strength of the
Mg\,{\sc i} line at 8807\,\AA\ for a subsample of stars from
CEN01a. Panel (a) shows stars with similar temperature and gravity but
spanning a wide range in metallicity. Panel (b) displays a sequence in
gravity for stars with similar temperature, and metallicity around
solar. Temperatures in K are given in brackets. All the spectra have
been normalized and reproduced using the same scale. It is clear the
increasing strength of the Mg\,{\sc i} line with the increasing
[Fe/H]. The effect of $\log g$ in the strength of the Mg\,{\sc i} line
is very mild, with intermediate $\log g$ stars having slightly weaker
lines.}
\label{secMgI}
\end{figure}

As it is expected for any metal line, Mg\,{\sc i} shows a negligible
strength in the spectra of the earliest spectral types (see the upper
spectrum in Figure~\ref{identl}) which are mainly dominated by strong
absorption Paschen lines. Therefore, high temperature stars have null
MgI values, or even below zero (Fig.~\ref{MgIsTiOtheta}a). The later
occurs since, in this kind of stars, the index pseudo-continuum lies
slightly below the true level thus leading to negative values when the
Mg\,{\sc i} line is very weak. For latter spectral types the Mg\,{\sc
i} strength increases with the decreasing temperature. This behaviour
peaks at $T_{\rm eff} \sim 4000 - 5000$\,K before decreasing for mid
and late-M types. For a wide range of spectral types ( $3500 \la
T_{\rm eff} \la 7000$\,K), the Mg\,{\sc i} line is also heavily
affected by metallicity and gravity effects leading to the spread of
MgI values in Fig.~\ref{MgIsTiOtheta}a.

The effects of metallicity and gravity on the Mg\,{\sc i} line are
also illustrated in Figure~\ref{secMgI}, where two comparative
sequences in (a) metallicity and (b) surface gravity for several G $-$
K spectral types from CEN01a around the the Mg\,{\sc i} line are
shown. From Figs.~\ref{MgIsTiOtheta}a and \ref{secMgI}a it is clear
that MgI increase as metallicity increases. On the contrary, the weak,
subtle dependence on gravity is difficult to distinguish at first
sight. Only when a detailed, statistical analysis is carryed out, it
is possible to detect that dwarfs and supergiants stars exhibit
slightly larger MgI indices that normal giants. The empirical fitting
functions derived in the next section will account for such a
behaviour.

\subsubsection{The TiO bands}
\label{qualstio}

\begin{figure}
\centerline{\hbox{
\psfig{figure=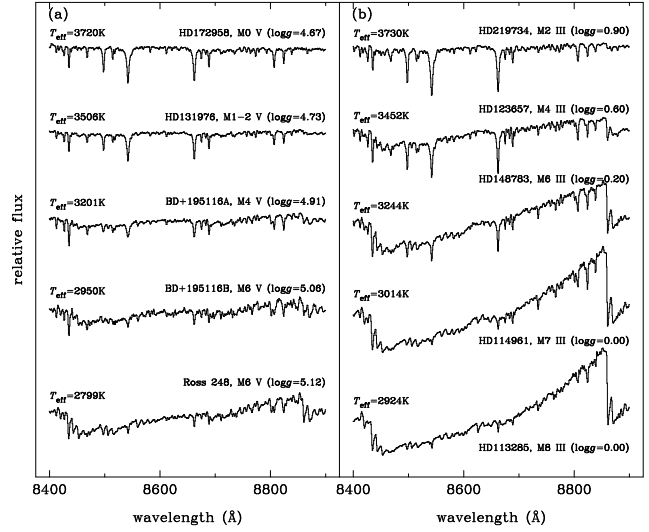,width=8.4cm,angle=-90}
}} 
\caption{\small Sequences in M-types for (a) dwarf and (b) giant stars
from CEN01a. Effective temperatures, names, spectral types, luminosity
classes and surface gravities ($\log g$ in dex) are given in the
labels. All the spectra have been normalized and reproduced using the
same scales so relative differences among the spectra are kept. At
fixed temperature, TiO bands in giants are stronger than in
dwarfs. For both luminosity classes, the strength of the TiO bands
increase with the decreasing temperature.}
\label{secTiO}
\end{figure}

As we already mentioned in Section~\ref{newindices}, molecular bands
of TiO and VO appear in the spectra of early M-types increasing their
strength with the decreasing temperature. Figure~\ref{secTiO} shows a
comparative sequence in late M-types for a sample of dwarfs (a) and
giants (b) from CEN01a. For a given temperature, giant stars exhibit
molecular bands quite stronger than those in dwarf stars. Also, the
strengthening rate of these bands with the decreasing temperature is
larger in giant stars. The last behaviours are apparent in
Fig.~\ref{MgIsTiOtheta}b, suggesting that such TiO bands can certainly
be used as a powerful dwarf-to-giant discriminator for cold spectral
types (see e.g. Gilbert et al.~2006). For $T_{\rm eff} \la 3600$\,K
($\theta \ga 1.4$\,K$^{-1}$), the index sTiO of dwarfs and giants
clearly follow two different, increasing trends. The rest of spectral
types exhibit sTiO $\la 1$. It makes sense since the local continuum
of their spectra is roughly flat. In spite of that, for this regime of
temperatures there exists a weak gravity dependence in the sense that
the lower the gravity the larger the index. It just arises from the
fact that the shape of the continuum slightly varies with the
luminosity class.

\subsection{The fitting functions}
\label{fitfunc}

In this section we present the empirical calibration of the new
indices in terms of the stellar atmospheric parameters. The outputs of
this procedure are the so called fitting functions, polynomials that
can be easily implemented into SSP codes to predict the integrated
indices of a wide variety of stellar systems (e.g.~Gorgas et al.~1993;
Worthey et al.~1994; Worthey \& Ottaviani 1997; Gorgas et al.~1999;
CEN02; Schiavon 2007; M\'armol-Queralt\'o et al.~2008; Maraston et
al.~2008).

The general procedure followed to compute the fitting functions is the
same as in CEN02, so we refer the reader to that paper for a detailed
description of the method. These fitting functions have been
calculated using the index measurements presented in
Section~\ref{newindexmeasurements} and the atmospheric parameters
derived in CEN01b. It is important to remind that the fitting
functions are only mathematical representations of the behaviour of
the indices as a function of the atmospheric parameters and, thus, a
physical justification of the derived coefficients is beyond the scope
of this paper.

Readers interested in employing these fitting functions can use the
{\sc fortran} routine available at \\{\tt
http://www.ucm.es/info/Astrof/ellipt/MgIsTiO.html}.\\ This program
performs the required interpolations to provide the MgI and sTiO
indices (together with CaT$^*$, CaT and PaT) as a function of the
three input atmospheric parameters. It also gives an estimation of the
errors in the index predictions, as it is explained in
Section~\ref{resanderr}.

\subsubsection{The fitting procedure}

Following the same procedure that in CEN02, we use $\theta$, $\log g$
and [Fe/H] as effective temperature, surface gravity and metallicity
indicators. The fitting functions have been computed as polynomials of
the atmospheric parameters with terms up to the third order, including
all possible cross--terms among the parameters. Two possible
functional forms are computed,

\begin{equation}
\label{ffpol}
{\cal I} _{\rm a}(\theta, \log g, {\rm [Fe/H]}) = p(\theta, \log g, {\rm [Fe/H]}), \;\;\;\; {\rm or} 
\end{equation}
\begin{equation}
\label{ffexp}
{\cal I} _{\rm a}(\theta, \log g, {\rm [Fe/H]}) = {\rm const.} + \exp{\left[p(\theta, \log g, {\rm [Fe/H]})\right]} , 
\end{equation}
keeping the one that minimizes the residuals of the fit. ${\cal I} _{\rm a}$ refers to
any of the above indices and $p$ is a polynomial 
\begin{equation}
p(\theta, \log g, {\rm [Fe/H]}) = \sum_{0 \le i+j+k \le 3}c_{i,j,k}\theta^{i}(\log g)^{j}{\rm [Fe/H]}^{k}, 
\label{eq3}
\end{equation}
with $0 \leq i+j+k \leq 3$  and $0 \leq i,j,k$.

Given the wide parameter space covered by the stellar sample and the
complex behaviour of the indices MgI and sTiO, we proceeded as in
CEN02 and divided the whole parameter space into several boxes of
parameters in which local fitting functions can be properly
computed. A final fitting function for the whole parameter space has
been constructed by interpolating the derived local functions. In
order to do that, the boundaries of the boxes were defined in such a
way that they overlapped, including thus several stars in common. In
the overlapping zones, cosine-weighted means of the functions
corresponding to both boxes were performed to guarantee a smooth
interpolation (see CEN02).

The local fitting functions were derived through a weighted least
squares fit to all the stars within each parameter box, with weights
according to the uncertainties of the indices for each individual
star, as given in Section~\ref{newindexmeasurements}. Since not all
the 20 possible terms were necessary, we followed a systematic
procedure to obtain the appropriate local fitting function in each
case. It made use of statistical criteria to accept or reject each
single term depending on its significance level (see CEN02 for further
details). Finally, the final combination of terms was the one which
provided the minimum unbiased residual variance.

\begin{table}                                                              
\centering{                                                                
\caption{List of stars which were rejected during the fitting function
  computation. The diagnostics for rejection are coded as: C, carbon
  star; EmL, emission lines of the elements in brackets; P, pulsating
  star; SB, spectroscopic binary; Var, variable star; *: bad quality
  spectrum because of low signal-to-noise ratios, bad exposing
  conditions, unreliable spectral features, and others.}
\label{rejected}                                                            
\begin{tabular}{@{}lll@{}ll@{}}          
\hline                                
Name & Diagnostic && Name & Diagnostic\\
\cline{1-2}
\cline{4-5}
HD 108   & EmL (Ca,H) &&      HD 112014  & SB           \\
HD 1326B & Flare star &&      HD 120933  & Var (CVn)    \\
HD 17491 & P          &&      HD 121447  & Var          \\
HD 35601 & P          &&      HD 181615  & EmL (Ca)     \\
HD 37160 & * MgI      &&      HD 217476  & Var          \\
HD 39801 & * MgI      &&      BD+ 61 154 & EmL (Ca,H)   \\
HD 42475 & P          &&   NGC 188 II-72 & * MgI        \\
HD 46687 & C          &&      M92 I-10   & * MgI        \\
HD 54300 & C          &&      M92 I-13   & * MgI \& sTiO\\
HD 58972 & SB         &&      M92 II-23  & * MgI \& sTiO\\
HD 74000 & * MgI      &&                 &              \\
\hline
\end{tabular}            
}                                                                          
\end{table}

During the fitting procedure for the MgI and sTiO indices, 21 and 16
stars were respectively rejected as they were found to exhibit large
residuals w.r.t.~the final fits, or were subject to have anomalous
behaviours. Such stars are indicated in Table~\ref{rejected}. Many of
them pose a variable, binary, or line-emiting nature, whilst some
other stars just have unreliable features affecting the index
measurement because of very low signal-to-noise ratios, cosmetic
defects, etc. For identical reasons, most of them were already
rejected for the final CaT$^{*}$, CaT, and PaT fitting functions in
CEN02.

The derived local fitting functions for the indices MgI and sTiO are
presented, respectively, in Tables~\ref{coeffsMgI} and
\ref{coeffssTiO}. The Tables are subdivided according to the
atmospheric parameters ranges of each fitting box and include: the
functional forms of the fits (polynomial or exponential), the
significant coefficients and their corresponding formal errors, the
typical index error for the $N$ stars employed in each interval
($\sigma^{2}_{\rm typ}=N/\sum_{i=1}^{N}\sigma_{i}^{-2}$), the unbiased
residual variance of the fit ($\sigma^{2}_{\rm std}$), and the
determination coefficient ($r^{2}$). Note that this coefficient
provides the fraction of the index variation in the sample which is
explained by the derived fitting functions. 

\begin{scriptsize}
\begin{table}
\centering{
\caption{Coefficients and statistical data of the local fitting
functions for the index MgI in each range of atmospheric
parameters. The term ``giant'' also includes supergiant stars.}
\label{coeffsMgI}
\vspace{4mm}
\begin{tabular}{@{}l@{}lr@{}c@{}ll@{}}          
\hline\hline    
(a) Hot dwarfs& & \multicolumn{3}{c}{0.13 $<$ $\theta$ $<$ 0.70} &
2.80 $<$ $\log$ $g$ $<$ 4.50 \\
\hline
exponential fit            & &    const. & ~ =  &~--1.50                           & $N$ = 49                              \\
$c_{0}$                    &:&   0.4284&~$\pm$&~0.0936                             & $\sigma_{{\rm typ}}$ = 0.046          \\
$\theta^{2}$               &:&  --4.833&~$\pm$&~1.137                              & $\sigma_{{\rm std}}$ = 0.094          \\
$\theta^{3}$               &:&    7.476&~$\pm$&~1.485                              & $r^{2}$ = 0.74                        \\
\hline\hline    
(b) Hot giants             & &   \multicolumn{3}{c}{0.13 $<$ $\theta$ $<$ 0.60}    & 1.20 $<$ $\log$ $g$ $<$ 3.01          \\
\hline
polynomial fit          & &         &     &                                     & $N$ = 21                              \\
$c_{0}$                    &:& --0.1202&~$\pm$&~0.0371                             & $\sigma_{{\rm typ}}$ = 0.033          \\
$\theta^{3}$               &:&   0.6254&~$\pm$&~0.2572                             & $\sigma_{{\rm std}}$ = 0.056          \\
                           & &         &     &                                     & $r^{2}$ = 0.47                        \\
\hline\hline    
(c) Warm stars             & &   \multicolumn{3}{c}{0.30 $<$ $\theta$ $<$ 1.30}    & 0.00 $<$ $\log$ $g$ $<$ 5.00          \\
\hline
polynomial fit          & &         &     &                                     & $N$ = 586                             \\
$c_{0}$                    &:&   1.846 &~$\pm$&~0.423                              & $\sigma_{{\rm typ}}$ = 0.044 \\
$\theta$                   &:& --7.960 &~$\pm$&~1.650                              & $\sigma_{{\rm std}}$ = 0.094          \\
$\log$ $g$                 &:&--0.1999 &~$\pm$&~0.0349                             & $r^{2}$ = 0.89                        \\
$\theta$~${\rm [Fe/H]}$    &:&  0.3079 &~$\pm$&~0.0310                             &                                       \\
$\theta^{2}$               &:&   11.68 &~$\pm$&~2.02                               &                                       \\
${\rm [Fe/H]}^{2}$         &:&  0.1862 &~$\pm$&~0.0600                             &                                       \\
$\theta^{2}$~$\log$ $g$    &:&--0.06817&~$\pm$&~0.03318                            &                                       \\
$\theta^{3}$               &:& --4.571 &~$\pm$&~0.760                              &                                       \\ 
$\theta$~${\rm [Fe/H]}^{2}$&:&--0.1624 &~$\pm$&~0.0605                             &                                       \\
$\theta$~$\log^{2}$$g$     &:&  0.05072&~$\pm$&~0.00568                            &                                       \\
\hline\hline    
(d) Cool stars             & &   \multicolumn{3}{c}{1.05 $<$ $\theta$ $<$ 1.35}    & 0.00 $<$ $\log$ $g$ $<$ 5.00          \\
\hline
polynomial fit          & &         &     &                                     & $N$ = 230                             \\
$c_{0}$                    &:&--9.005  &~$\pm$&~3.563                              & $\sigma_{{\rm typ}}$ = 0.039 \\   
$\theta$                   &:&  16.76  &~$\pm$&~6.00                               & $\sigma_{{\rm std}}$ = 0.099          \\
$\theta$~${\rm [Fe/H]}$    &:&  1.711  &~$\pm$&~0.641                              & $r^{2}$ = 0.76                        \\
$\theta$~$\log$$g$         &:&--0.1262 &~$\pm$&~0.0400                             &                                       \\
$\theta^{2}$               &:&--6.941  &~$\pm$&~2.516                              &                                       \\
${\rm [Fe/H]}^{2}$         &:&  0.8445 &~$\pm$&~0.4527                             &                                       \\
$\log^{3}$$g$              &:&0.007011 &~$\pm$&~0.001746                           &                                       \\
$\theta^{2}$~${\rm [Fe/H]}$&:&--1.247  &~$\pm$&~0.536                              &                                       \\
$\theta$~${\rm [Fe/H]}^{2}$&:&--0.7490 &~$\pm$&~0.3924                             &                                       \\
\hline\hline    
(e) Cold dwarfs            & &   \multicolumn{3}{c}{1.07 $<$ $\theta$ $<$ 1.90}    & 4.40 $<$ $\log$ $g$ $<$ 5.20          \\
\hline
polynomial fit          & &         &     &                                     & $N$ = 22                              \\
$c_{0}$                    &:&   2.638 &~$\pm$&~0.469                              & $\sigma_{{\rm typ}}$ = 0.039          \\
$\theta^{2}$               &:& --2.098 &~$\pm$&~0.678                              & $\sigma_{{\rm std}}$ = 0.085          \\
$\theta^{3}$               &:&   0.805 &~$\pm$&~0.306                              & $r^{2}$ = 0.90                        \\
\hline\hline    
(f) Cold giants            & &   \multicolumn{3}{c}{1.30 $<$ $\theta$ $<$ 1.80}    & 0.00 $<$ $\log$ $g$ $<$ 1.65          \\
\hline
polynomial fit          & &         &     &                                     & $N$ = 27                              \\
$c_{0}$                    &:&   1.896 &~$\pm$&~0.254                              & $\sigma_{{\rm typ}}$ = 0.023 \\
$\theta^{2}$               &:&--0.5762 &~$\pm$&~0.1161                             & $\sigma_{{\rm std}}$ = 0.157          \\
                           & &         &      &                                    & $r^{2}$ = 0.74                        \\
\hline\hline
\end{tabular}                                                                   
}
\end{table}   

\begin{table}
\centering{
\caption{Coefficients and statistical data of the local fitting
functions for the index sTiO in each range of atmospheric
parameters. The term ``giant'' also includes supergiant stars.}
\label{coeffssTiO}
\vspace{4mm}
\begin{tabular}{@{}l@{}lr@{}c@{}ll@{}}          
\hline\hline    
(a) Hot dwarfs      & & \multicolumn{3}{c}{0.13 $<$ $\theta$ $<$ 0.65} &
2.81 $<$ $\log$ $g$ $<$ 4.37 \\
\hline
exponential fit            & &  const. & ~ =  &~0.78                               & $N$ = 35                              \\
$c_{0}$                    &:&  --2.261&~$\pm$&~0.059                              & $\sigma_{{\rm typ}}$ = 0.004          \\
$\theta^{2}$               &:&    8.062&~$\pm$&~0.916                              & $\sigma_{{\rm std}}$ = 0.013          \\
$\theta^{3}$               &:&  --11.07&~$\pm$&~1.31                               & $r^{2}$ = 0.78                        \\
\hline\hline    
(b) Hot giants             & &   \multicolumn{3}{c}{0.13 $<$ $\theta$ $<$ 0.85}    & 0.39 $<$ $\log$ $g$ $<$ 3.01          \\
\hline
polynomial fit          & &         &     &                                     & $N$ = 40                              \\
$c_{0}$                    &:&   0.9551&~$\pm$&~0.0066                             & $\sigma_{{\rm typ}}$ = 0.003          \\
                           & &         &      &                                    & $\sigma_{{\rm std}}$ = 0.030          \\
                           & &         &     &                                     & $r^{2}$ = 0.61                        \\
\hline\hline    
(c) Warm stars             & &   \multicolumn{3}{c}{0.60 $<$ $\theta$ $<$ 1.30}    & 0.00 $<$ $\log$ $g$ $<$ 4.85         \\
\hline
polynomial fit          & &         &     &                                     & $N$ = 569                             \\
$c_{0}$                    &:&   1.855 &~$\pm$&~0.223                              & $\sigma_{{\rm typ}}$ = 0.004 \\
$\theta$                   &:& --2.618 &~$\pm$&~0.700                              & $\sigma_{{\rm std}}$ = 0.015          \\
$\log$ $g$                 &:&--0.07343&~$\pm$&~0.01508                            & $r^{2}$ = 0.75                        \\
${\rm [Fe/H]}$             &:&  0.03046&~$\pm$&~0.00511                            &                                       \\
$\theta^{2}$               &:&   2.691 &~$\pm$&~0.726                              &                                       \\
$\log^{2}$$g$              &:&  0.01767&~$\pm$&~0.00661                            &                                       \\
$\log$ $g$~${\rm [Fe/H]}$  &:&--0.008611&~$\pm$&~0.001667                          &                                       \\
$\theta^{3}$               &:& --0.8813&~$\pm$&~0.2471                             &                                       \\ 
$\log^{3}$$g$              &:&--0.001651&~$\pm$&~0.000857                          &                                       \\
\hline\hline    
(d) Cold dwarfs            & &   \multicolumn{3}{c}{1.07 $<$ $\theta$ $<$ 1.90}    & 4.45 $<$ $\log$ $g$ $<$ 5.13          \\
\hline
polynomial fit          & &         &     &                                     & $N$ = 21                              \\
$c_{0}$                    &:&  2.454  &~$\pm$&~0.296                              & $\sigma_{{\rm typ}}$ = 0.004 \\   
$\theta$                   &:&--2.547  &~$\pm$&~0.418                              & $\sigma_{{\rm std}}$ = 0.017          \\
$\theta^{2}$               &:&  1.070  &~$\pm$&~0.145                              & $r^{2}$ = 0.98                        \\
\hline\hline    
(e) Cold giants            & &   \multicolumn{3}{c}{1.28 $<$ $\theta$ $<$ 1.47}    & 0.00 $<$ $\log$ $g$ $<$ 2.00         \\
\hline
polynomial fit          & &         &     &                                     & $N$ = 25                              \\
$c_{0}$                    &:&   5.997 &~$\pm$&~1.925                              & $\sigma_{{\rm typ}}$ = 0.003 \\
$\theta^{2}$               &:& --9.245 &~$\pm$&~3.119                              & $\sigma_{{\rm std}}$ = 0.024          \\
$\theta^{3}$               &:&   4.837 &~$\pm$&~1.524                              & $r^{2}$ = 0.95                        \\
\hline\hline    
(f) Very cold giants       & &   \multicolumn{3}{c}{1.40 $<$ $\theta$ $<$ 1.80}    & 0.00 $<$ $\log$ $g$ $<$ 1.65          \\
\hline
polynomial fit          & &         &     &                                     & $N$ = 14                              \\
$c_{0}$                    &:& --39.79 &~$\pm$&~5.45                               & $\sigma_{{\rm typ}}$ = 0.006 \\
$\theta$                   &:&   37.44 &~$\pm$&~5.24                               & $\sigma_{{\rm std}}$ = 0.006          \\
$\theta^{3}$               &:& --4.319 &~$\pm$&~0.709                              & $r^{2}$ = 0.99                        \\
\hline\hline
\end{tabular}                                                                   
}
\end{table}   
\end{scriptsize}

\subsubsection{MgI fitting functions}
\label{mgiffs}

\begin{figure*}
\centerline{\hbox{
\psfig{figure=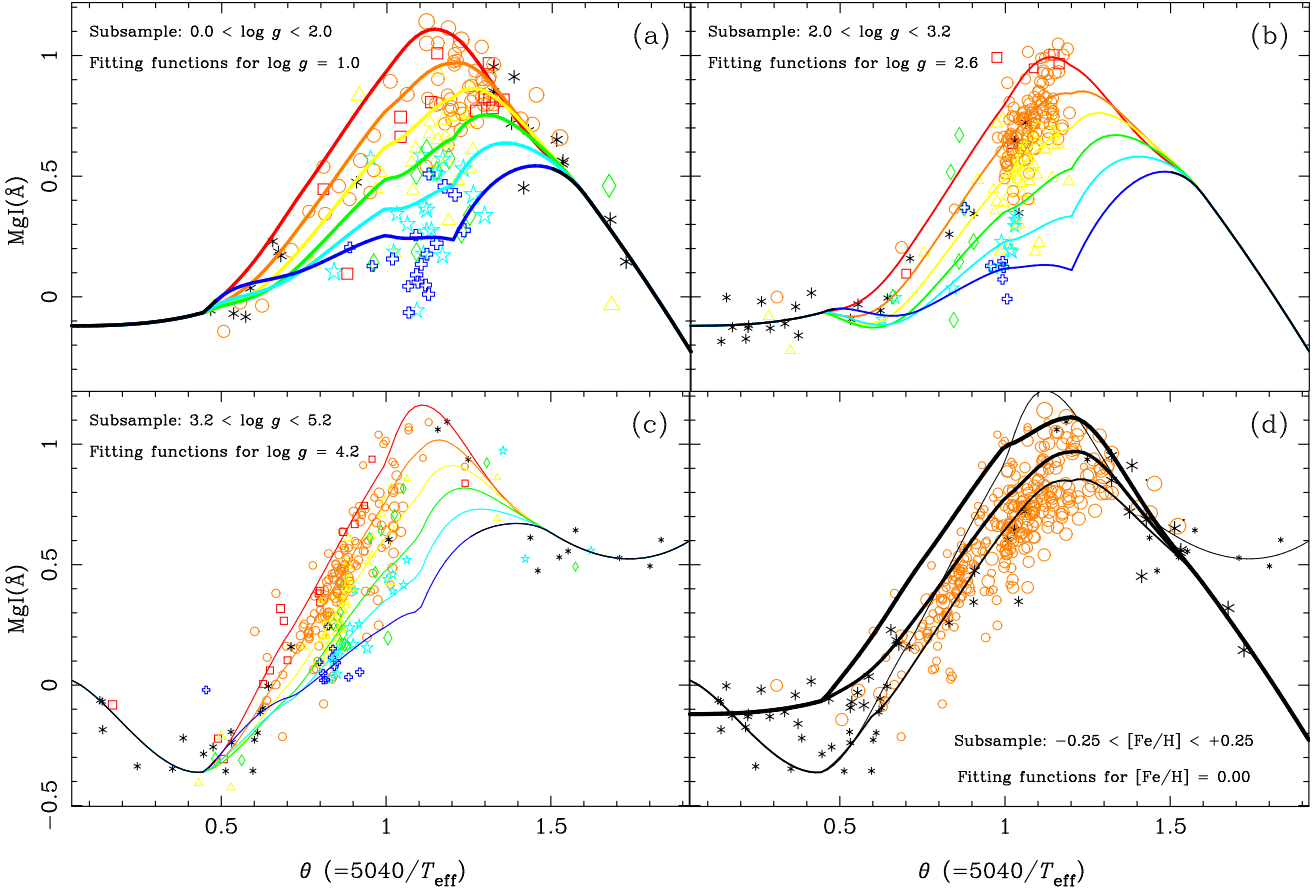}
}}
\caption{MgI values and general fitting functions for different
  atmospheric parameter regimes. Panels $a$, $b$ and $c$ display,
  respectively, all the stars with gravities in the ranges
  $0.0\leq\log g<2.0$, $2.0\leq\log g<3.2$ and $3.2\leq\log g<5.2$,
  together with the derived fitting functions for the mean gravity in
  each range, i.e., $\log g =$ 1.0 (a), 2.6 (b) and 4.2 (c). In the
  mid-temperature range, the different lines represent, from top to
  bottom, the functions for metallicities [Fe/H] = $+0.5$, 0.0,
  $-0.5$, $-1.0$, $-1.5$, and $-2.0$, whilst, for high and low
  temperatures, the fitting functions do not depend on
  metallicity. Panel $d$ shows all the stars around solar metallicity
  ($-0.25<{\rm [Fe/H]}\leq+0.25$) and the corresponding fitting
  functions computed for [Fe/H] = 0.0 and different values of gravity
  ($\log g$ = 0.0, 1.0, 3.0, and 5.0, from the thickest to the
  thinnest line). Codes and relative sizes of the star symbols
  (indicating, respectively, metallicity and gravity ranges) are
  explained in Fig.~\ref{MgIsTiOtheta}a. Note that, while the lines
  displayed here correspond to fitting functions at particular values
  of $\log g$ and [Fe/H], the plotted stars span a range of
  atmospheric parameters around these central values. This is the
  reason why the lines do not exactly fit all the points in the
  plots. Also, note that these fitting functions have not been derived
  by using only these plotted stars, but the whole sample.}
\label{fitMgI}
\end{figure*}

At both ends of the temperature regime covered by our stellar library,
MgI exhibits a different behaviour for dwarf stars, on one side, and
giant and supergiants stars, on the other one. This is why we defined
separate boxes in Table~\ref{coeffsMgI} to compute the corresponding
local fits: $a$ and $b$, for hot dwarfs and giants; $e$ and $f$, for
cold dwarfs and giants. No terms in $\log g$ and [Fe/H] were found to
be statistically significant in any of the two luminosity bins, so
only terms in $\theta$ were needed. However, for intermediate
temperatures (warm and cool stars; boxes $c$ and $d$ in
Table~\ref{coeffsMgI}), the three atmospheric parameters play an
important role to reproduce the complex behaviour of the MgI index.

Figures~\ref{fitMgI}a, \ref{fitMgI}b, and \ref{fitMgI}c illustrate the
general fitting functions (that is, the interpolation of all the local
fitting functions) for three gravity bins, that typically represent
supergiants, giants, and dwarfs, respectively. As expected, there
exists a clear dependence on metallicity and $\theta$ at intermediate
temperatures in the sense that MgI increases with the increasing
[Fe/H] and the decreasing temperature. Figure~\ref{fitMgI}d displays
the derived fitting functions for solar metallicity and different
gravity values. Apart from reinforcing the strong temperature
dependence of the MgI index, this figure illustrates that gravity
effects are not negligible at intermediate temperatures. As a matter
of fact, there exists a sort of degeneracy in the sense that dwarf
stars (thinest solid lines; $\log g \sim 4 - 5$) and supergiant stars
(thickest solid lines; $\log g \sim 0 - 1$) reach similar MgI values,
the ones are systematically larger than those of giants with
intermediate $\log g$ values.

It is important to note that, while the lines in Figure~\ref{fitMgI}
correspond to projections of the fitting functions at particular
values of $\log g$ and [Fe/H], the plotted stars span a range of
atmospheric parameters around these central values. Therefore, it is
not expected the lines to fit exactly all the points in the plots. In
any case, there are still some curves which do not appear to be well
constrained by the observations in certain regions of the parameter
space. For instance, there are no stars for $\theta < 0.5$ in
Figure~\ref{fitMgI}a ($\log g = 1.0$) and for $\theta > 1.3$ in
Figure~\ref{fitMgI}b ($\log g = 2.6$). Stars with these parameters
just do not exist, and therefore they will never be required by the
stellar populations models.

Because of the complicated functional form of MgI at intermediate
temperatures as compared to that of cold stars, the local fit in
Table~\ref{coeffsMgI}d was specially designed to preserve smoothness
when interpolating the local fits of warm and cold stars. Even so, a
ficticious peak in the fitting functions of very metal-poor ([Fe/H]
$\sim -2.0$) giants and supergiants at $\theta \sim 1.2$ is still
apparent in Figures~\ref{fitMgI}a,b. This is in part due to the lack
of reliable metallicity determinations for very cold stars, which
prevents us from a more accurate calibration of metallicity effects
all over the $\theta$--$\log g$ space. As a matter of fact, there
exists a stronger limitation arising from the intrinsic absence of
metal-poor, very cold giant stars, as the red giant and asymptotic
giant branches of so metal-poor populations ---like, e.g., the
globular cluster M92--- do not reach so low temperatures. Once again,
we are confident that the above interpolation is not having an
important effect on the integrated MgI values computed for
low-metallicity SSPs.

It is worth noting here that, aside from studying the behaviour of the
Mg\,{\sc i} line, DTT also calibrated the strength of the MgI(DTT)
index as a function of the stellar atmospheric parameters. By
performing a principal component analysis for the 106 stars of their
library (F5 to M1 spectral types), they derived a biparametrical
linear dependence on metallicity and effective temperature ---with no
sensitivity to surface gravity---, so that MgI(DTT) increases with the
increasing [Fe/H] and the decreasing $T_{\rm eff}$. In spite of the
parameter coverage of the DTT stellar sample allowing to detect the
weak dependence on $\log g$ and other significant terms, they
restricted their analysis to the two first principal components, what
explains the simple dependence reported in their work. For this
reason, the DTT calibration, although useful for achieving a first
order understanding of the MgI behaviour, should not be considered as
an accurate input ingredient for SSP modeling. As a matter of fact, a
simple extrapolation of the DTT calibration to larger and lower
temperatures fails to reproduce, among other, the observed MgI
turnover at $\theta \sim 1.2$ (after which MgI decreases with the
decreasing $T_{\rm eff}$) and the MgI plateu for the earliest spectral
types.

\subsubsection{sTiO fitting functions}
\label{stioffs}

\begin{figure}
\centerline{\hbox{
\psfig{figure=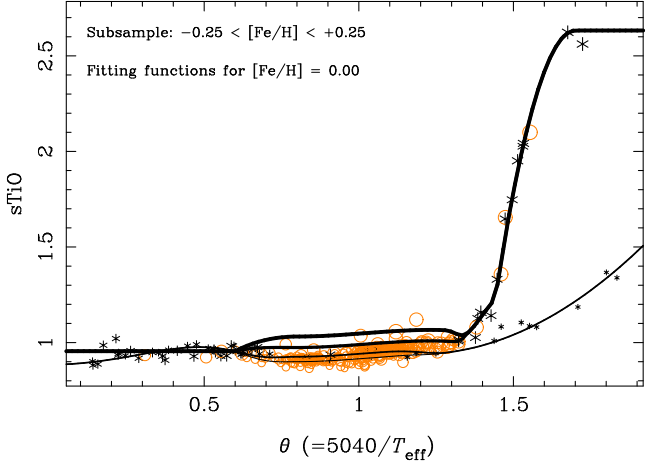,width=8.5cm,angle=-90}
}}
\caption{sTiO values for all the stars around solar metallicity
($-0.25<{\rm [Fe/H]}\leq+0.25$). The curves correspond to general
fitting functions computed for [Fe/H] = 0.0 and different values of
gravity ($\log g$ = 0.0, 1.0, 3.0, and 5.0, from the thickest
to the thinnest line). Codes and relative sizes of the star symbols
(indicating, respectively, metallicity and gravity ranges) are
explained in Fig.~\ref{MgIsTiOtheta}a.}

\label{fitsTiO}
\end{figure}

As it was described in Section~\ref{qualstio}, the index sTiO exhibits
values around 1 ---or slightly smaller--- for most library stars except
for the latest spectral types, for which the index increases with the
decreasing temperature. Boxes $d$, $e$, and $f$ in
Table~\ref{coeffssTiO} were designed to reproduce, separately for
dwarf and giant stars, the increasing sTiO trend at the low
temperature regime. Only terms in $\theta$ were necessary, as it also
happens for the earliest spectral types (boxes $a$ and $b$).

In spite of the apparent constancy of the sTiO values for intermediate
temperatures, terms in all the three atmospheric parameters turned out
to be statistically significant in box $c$, with the [Fe/H] dependence
being the less important. For this reason, Figure~\ref{fitsTiO} ---for
solar metallicity and different gravities--- suffices to illustrate the
general fitting functions of the sTiO index. Apart from the strong
sTiO increase with the decreasing temperature (particularly for cold
giant stars), it is worth noting the gravity effect at the
intermediate temperature regime, in the sense that the larger the
gravity (dwarfs; thinest solid lines), the lower the sTiO index.

\subsection{Residuals and error analysis}
\label{resanderr}

\begin{figure}
\centerline{\hbox{
\psfig{figure=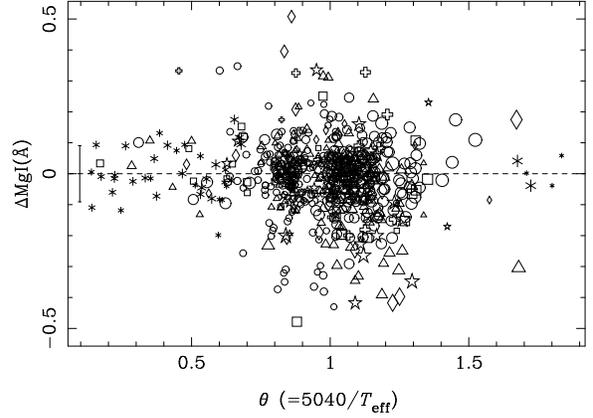,width=3.0in,angle=-90}
}}
\caption{Residuals of the MgI fitting functions ($\Delta {\rm MgI} =
    {\rm MgI}_{\rm obs} - {\rm MgI}_{\rm pred}$) versus $\theta$ for
    the whole stellar library. See Fig.~\ref{MgIsTiOtheta}a for symbol
    codes. The error bar at the left margin indicates the unbiased
    residual standard deviation of the fit. See text in
    Sections~\ref{resanderr} and ~\ref{MgFeratio} for a study on the
    different error sources driving the observed residuals.}
\label{residualsMgI}
\end{figure}

\begin{figure}
\centerline{\hbox{
\psfig{figure=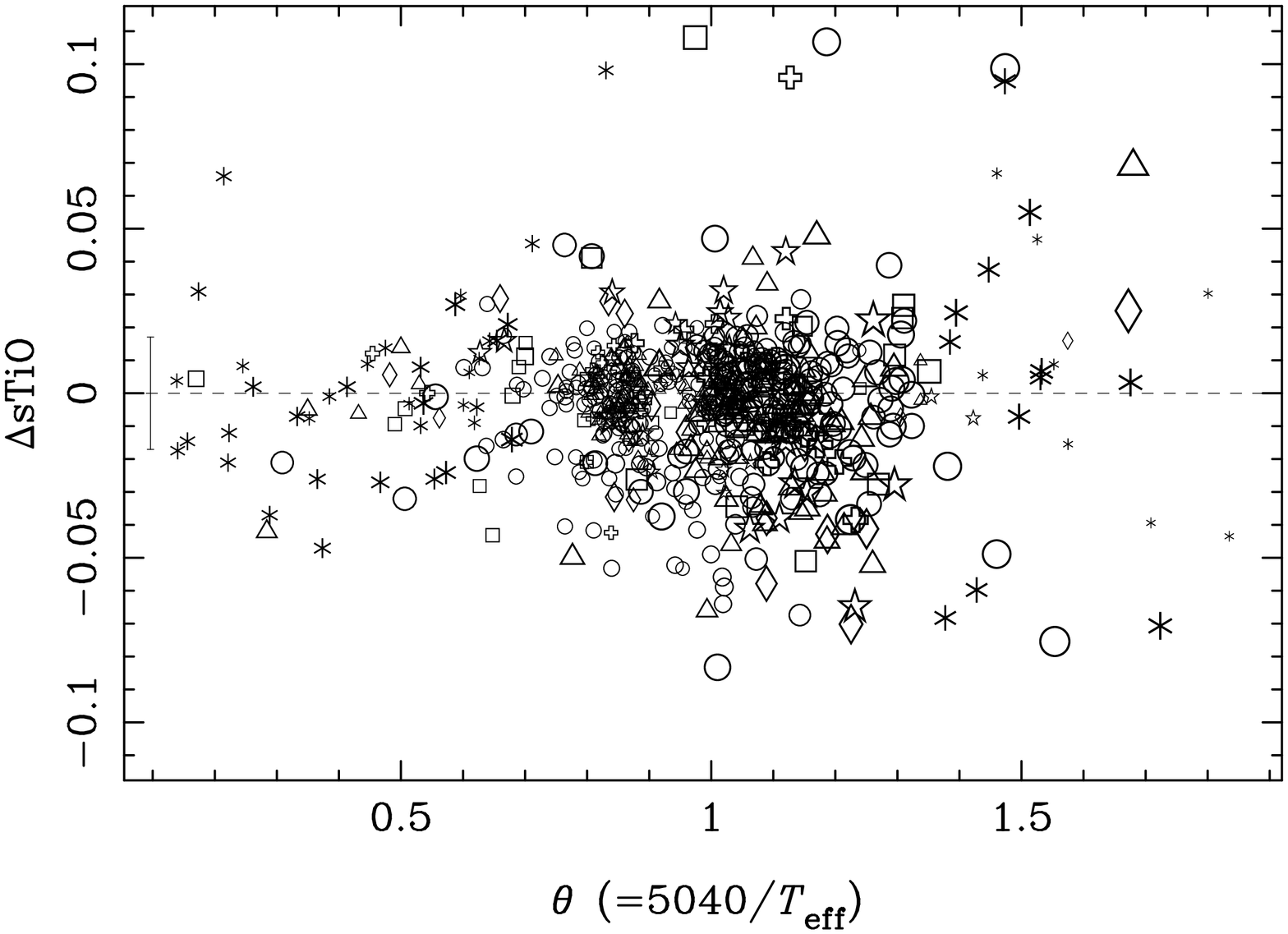,width=3.0in,angle=-90}
}}
\caption{Residuals of the sTiO fitting functions ($\Delta {\rm sTiO} =
  {\rm sTiO}_{\rm obs} - {\rm sTiO}_{\rm pred}$) versus $\theta$ for
  the whole stellar library. See Fig.~\ref{MgIsTiOtheta}a for symbol
  codes. The error bar at the left margin indicates the unbiased
  residual standard deviation of the fit. See text in
  Section~\ref{resanderr} for a study on the different error sources
  driving the observed residuals.}
\label{residualssTiO}
\end{figure}

Defining the index residual of a given star as the difference between
the observed index and the one predicted by the fitting functions
($\Delta {\cal I} = {\cal I}_{\rm obs} - {\cal I}_{\rm pred}$),
Figures~\ref{residualsMgI} and \ref{residualssTiO} show the residuals
of the indices MgI and sTiO for the whole stellar library as a
function of $\theta$. These residuals are given for each star in
Table~\ref{macrotab}. Overall, no systematic deviations are found for
any of the three atmospheric parameters. Star clusters have also been
analyzed separately and, except for the MgI values of the open cluster
M67, no systematic effects have been found. We defer the analysis of
this particular case to Section~\ref{MgFeratio}, where the effect of
different [Mg/Fe] ratios on the MgI fitting function residuals is
discussed.

\begin{table}                         
\centering{                                                                
\caption{Statistical data for the general fitting functions of the
indices MgI and sTiO. N: number of stars; $\sigma_{\rm std}$: unbiased
residual standard deviation; $\sigma_{\rm typ}$: typical index error
for the stars used in the fits; $r^{2}$: determination coefficient.}
\begin{tabular}{lcccc}
\hline
        & N & $\sigma_{\rm std}$ & $\sigma_{\rm typ}$ & $r^{2}$ \\
\hline	  	  	  
MgI     &647& 0.091               & 0.043               & 0.91    \\
sTiO    &668& 0.017               & 0.004               & 0.97    \\
\hline
\end{tabular}
\label{sigmas}
}
\end{table}

\begin{table*}                        
\centering{                                                                
\caption{Uncertainties of the MgI and sTiO fitting functions for
  different subsets of stars, and mean index errors due to
  uncertainties in the flux calibration and the input atmospheric
  parameters. Open clusters are Coma, Hyades, M67, NGC188 and
  NGC7789. Globular clusters include M3, M5, M10, M13, M71, M92 and
  NGC6171. $N$: number of stars. $\sigma_{\rm typ}$: typical
  observational index error, accounting for photon noise and typical
  uncertainties in radial velocity. $\sigma_{T_{\rm eff}}$, $\sigma_{\log g}$ and
  $\sigma_{\rm [Fe/H]}$: mean index errors due to uncertainties in the
  input $T_{\rm eff}$, $\log g$ and [Fe/H]. $\sigma_{{\rm par}}$:
  total error due to atmospheric parameters (quadratic addition of the
  three previous ones). $\sigma_{\rm fcal}$: mean index
  error arising from random uncertainties in the flux
  calibration. $\sigma_{{\rm ALL}}$: total error due to all
  the above uncertainties (quadratic addition of $\sigma_{\rm typ}$,
  $\sigma_{{\rm par}}$, and $\sigma_{\rm fcal}$). $\sigma_{\rm std}$:
  unbiased residual standard deviation of the fit.}
\begin{tabular}{@{}lcrcccccccccccc@{}}
\hline
       & Index & $N$ & &  $\sigma_{\rm typ}$ & & $\sigma_{T_{\rm eff}}$ & $\sigma_{\log g}$ & $\sigma_{\rm [Fe/H]}$ & $\sigma_{{\rm par}}$ & & $\sigma_{\rm fcal}$ & & $\sigma_{\rm ALL}$ & $\sigma_{\rm std}$ \\
\hline
Open clusters                              & MgI  &  92 & & 0.065 & & 0.037 &0.009 &0.056 &0.068 & & 0.007 & & 0.094 & 0.114 \\
                                           & sTiO &  93 & & 0.006 & & 0.003 &0.001 &0.001 &0.003 & & 0.011 & & 0.013 & 0.017 \\ 
Globular clusters                          & MgI  &  53 & & 0.174 & & 0.019 &0.006 &0.056 &0.060 & & 0.007 & & 0.184 & 0.219 \\
                                           & sTiO &  54 & & 0.015 & & 0.002 &0.002 &0.003 &0.004 & & 0.011 & & 0.019 & 0.028 \\
Field dwarfs                               & MgI  & 236 & & 0.042 & & 0.030 &0.029 &0.026 &0.049 & & 0.007 & & 0.065 & 0.082 \\
                                           & sTiO & 242 & & 0.004 & & 0.004 &0.003 &0.001 &0.005 & & 0.012 & & 0.014 & 0.011 \\
Field giants                               & MgI  & 196 & & 0.036 & & 0.026 &0.009 &0.028 &0.039 & & 0.007 & & 0.054 & 0.091 \\
                                           & sTiO & 204 & & 0.003 & & 0.017 &0.004 &0.001 &0.017 & & 0.012 & & 0.021 & 0.016 \\
Field supergiants                          & MgI  &  70 & & 0.039 & & 0.030 &0.025 &0.030 &0.049 & & 0.007 & & 0.063 & 0.092 \\
                                           & sTiO &  75 & & 0.004 & & 0.023 &0.007 &0.002 &0.024 & & 0.012 & & 0.027 & 0.031 \\
\hline		     			            	      	                	                                   
Hot stars ($0.13 < \theta < 0.69$)         & MgI  &  67 & & 0.040 & & 0.054 &0.014 &0.023 &0.061 & & 0.007 & & 0.073 & 0.086 \\
                                           & sTiO &  68 & & 0.004 & & 0.010 &0.003 &0.000 &0.011 & & 0.011 & & 0.016 & 0.022 \\
Intermediate stars ($0.69 < \theta < 1.30$)& MgI  & 555 & & 0.044 & & 0.023 &0.018 &0.036 &0.047 & & 0.007 & & 0.065 & 0.090 \\
                                           & sTiO & 560 & & 0.004 & & 0.001 &0.003 &0.001 &0.004 & & 0.012 & & 0.013 & 0.016 \\
Cold stars ($1.30 < \theta < 1.84$)        & MgI  &  25 & & 0.029 & & 0.061 &0.009 &0.006 &0.062 & & 0.007 & & 0.069 & 0.106 \\
                                           & sTiO &  40 & & 0.004 & & 0.123 &0.004 &0.002 &0.123 & & 0.022 & & 0.125 & 0.025 \\
\hline		     			   		                         	                                   
All                                        & MgI  & 647 & & 0.043 & & 0.029 &0.018 &0.034 &0.048 & & 0.007 & & 0.065 & 0.091 \\
                                           & sTiO & 668 & & 0.004 & & 0.009 &0.003 &0.001 &0.010 & & 0.012 & & 0.016 & 0.017 \\
\hline
\end{tabular}
\label{errpredatm}
}
\end{table*}

To explore in more detail the reliability of the present fitting
functions, in Table~\ref{sigmas} we list the unbiased residual
standard deviation from the fits, $\sigma_{\rm std}$, the typical
error in the measured indices arising from photon noise and radial
velocity uncertainties, $\sigma_{\rm typ}$, and the determination
coefficient, $r^{2}$, for all the stars employed in the computation of
the general fitting functions. It is important to note that, apart
from the outlier stars rejected from the fits, a few stars with
unknown [Fe/H] could not be included in the metallicity-dependent fits
and no residuals were therefore computed. It is clear that, for both
indices, $\sigma_{\rm std}$ is larger than what it should be expected
uniquely from typical errors (see also the partial values of
$\sigma_{\rm std}$ and $\sigma_{\rm typ}$ in Tables~\ref{coeffsMgI}
and~\ref{coeffssTiO}), what suggests that the fitting function
residuals must be dominated by other effects. In CEN02 it was
demonstrated that the errors in the input parameters were the main
source of residuals for the fitting functions of the Ca\,{\sc ii}
indices. We therefore perform a similar analysis to constrain the
effect of the atmospheric parameter uncertainties in the residuals of
the MgI and sTiO fitting functions. Aimed at constraining the
potential sensitivity of sTiO to small changes in the continuum shape,
the effect of flux calibration uncertainties is also discussed.

\subsubsection{Uncertainties in the stellar atmospheric parameters}
\label{uncatmpar}

We have computed how the errors in the input atmospheric parameter of
the library stars translate into uncertainties in the predicted
indices. Since this depends on both the local functional form of the
fitting functions (e.g., a weak dependence on temperature leads to
small index errors due to $T_{\rm eff}$ uncertainties) and the
atmospheric parameters range (e.g., both hot and very cold stars have
$T_{\rm eff}$ uncertainties larger than intermediate temperature
stars), we have not only performed the analysis for the stellar
library as a whole but also for different sets of stars (listed in
Table~\ref{errpredatm}).

For each star of the sample we have derived three index errors,
arising from the corresponding uncertainties in $T_{\rm eff}$, $\log
g$ and [Fe/H]. As input atmospheric parameters uncertainties we have
made use of the values presented in Table~7 of CEN01b. Apart from
those, we have used errors of 75\,K, 0.40\,dex and 0.15\,dex for the
effective temperatures, gravities and metallicities taken from
Soubiran, Katz \& Cayrel (1998) (with 4000\,K $<T_{\rm eff}<$
6300\,K; stars coded {\sc skc} in Table~6 of CEN01b), and 75\,K,
0.05\,dex and 0.20\,dex for the cluster stars.  For each subset of
stars we have computed a mean index error as a result of the
uncertainty of each parameter ($\sigma_{T_{\rm eff}}$, $\sigma_{\log
g}$ and $\sigma_{\rm [Fe/H]}$) by using the input parameter errors for
all the individual stars. Finally, an estimate of the total expected
error due to atmospheric parameters ($\sigma_{{\rm par}}$) is computed
as the quadratic addition of the three previous errors.

It is clear from the data in Table~\ref{errpredatm} that $\sigma_{{\rm
par}}$ is, in all cases, comparable or larger than $\sigma_{{\rm
typ}}$. This result reinforces the importance of using an homogeneous
and reliable set of atmospheric parameters to guarantee the accuracy
of this kind of calibrations. Also, it is interesting to see how
$\sigma_{T_{\rm eff}}$ for cold stars is much larger than their
observed dispersion w.r.t. the fits, $\sigma_{\rm std}$. This would
mean that the error in $T_{\rm eff}$ quoted in CEN01b for this types
of stars was somewhat overestimated.

\begin{table}                         
\centering{                                                                
\caption{Absolute errors in the fitting functions predictions for
different values of the atmospheric parameters. Input $\log g$ values
varying with effective temperature for dwarfs, giants and supergiants
have been taken from Lang (1991). Since, for extreme temperatures, the
fitting functions do not depend on metallicity, no [Fe/H] value has
been adopted for 15000\,K and 3200\,K. This is also the case for some
values at 3500\,K.}
\begin{tabular}{@{}r@{}rc@{}cc@{}cc@{}c@{}}
\hline
    & &\multicolumn{2}{c}{dwarfs}&\multicolumn{2}{c}{giants}&\multicolumn{2}{c}{supergiants}\\
\hline
$T_{\rm eff}$\ \ &[Fe/H]& $\Delta$sTiO\ \ &$\Delta$MgI\ \ & $\Delta$sTiO\ \ &$\Delta$MgI\ \ & $\Delta$sTiO\ \ &$\Delta$MgI\\
\hline \medskip
15000&      & 0.001 & 0.041 & 0.001 & 0.041 & 0.007 & 0.029\\
 8000&$+$0.5& 0.002 & 0.031 & 0.002 & 0.031 & 0.007 & 0.039\\
 8000&   0.0& 0.002 & 0.029 & 0.002 & 0.031 & 0.007 & 0.039\\
 8000&$-$1.0& 0.003 & 0.034 & 0.002 & 0.034 & 0.007 & 0.042\\ \medskip
 8000&$-$2.0& 0.003 & 0.082 & 0.003 & 0.080 & 0.007 & 0.082\\
 6000&$+$0.5& 0.003 & 0.023 & 0.004 & 0.023 & 0.006 & 0.033\\
 6000&   0.0& 0.002 & 0.013 & 0.003 & 0.016 & 0.006 & 0.029\\
 6000&$-$1.0& 0.003 & 0.018 & 0.004 & 0.021 & 0.006 & 0.033\\ \medskip
 6000&$-$2.0& 0.006 & 0.048 & 0.005 & 0.048 & 0.007 & 0.054\\
 5000&$+$0.5& 0.005 & 0.045 & 0.003 & 0.031 & 0.006 & 0.038\\
 5000&   0.0& 0.004 & 0.031 & 0.002 & 0.016 & 0.005 & 0.028\\
 5000&$-$1.0& 0.004 & 0.032 & 0.003 & 0.022 & 0.005 & 0.029\\ \medskip
 5000&$-$2.0& 0.007 & 0.048 & 0.005 & 0.038 & 0.007 & 0.039\\
 4000&$+$0.5& 0.006 & 0.063 & 0.005 & 0.064 & 0.009 & 0.078\\
 4000&   0.0& 0.006 & 0.051 & 0.004 & 0.037 & 0.008 & 0.063\\
 4000&$-$1.0& 0.006 & 0.052 & 0.004 & 0.055 & 0.007 & 0.066\\ \medskip
 4000&$-$2.0& 0.007 & 0.084 & 0.007 & 0.120 & 0.009 & 0.114\\
 3500&$+$0.5& 0.010 & 0.045 & 0.027 & 0.072 & 0.027 & 0.073\\
 3500&   0.0& 0.010 & 0.044 & 0.027 & 0.067 & 0.027 & 0.068\\
 3500&$-$1.0& 0.010 & 0.046 & 0.027 & 0.080 & 0.027 & 0.081\\ \medskip
 3500&$-$2.0& 0.010 & 0.050 & 0.027 & 0.104 & 0.027 & 0.103\\
 3200&      & 0.011 & 0.041 & 0.042 & 0.063 & 0.042 & 0.063\\
\hline
\end{tabular}
\label{errpred}
}
\end{table}

Furthermore, since the aim of this paper is to predict reliable index
values for any combination of input atmospheric parameters, we have
also computed the random errors in such predictions making use of the
covariance matrices of the fits. These uncertainties are given in
Table~\ref{errpred} for some representative values of input
parameters. Note that, as it is expected, the absolute errors are
larger for cold giants and supergiants and, in general, increase as
the metallicity departs from the solar value. These errors are the
ones provided by the {\sc fortran} routine that computes the fitting
function predictions as uncertainties of the output MgI and sTiO
indices.

\subsubsection{Uncertainties in the flux calibration }
\label{uncfluxcal}

An additional source of index errors that may increase the fitting
function residuals is the uncertainty in the flux calibration of the
library stars. At this point, we are not interested in the quality of
the flux calibration in an absolute sense, as any minor departure from
the ``true'' calibration must be considered as a systematic that
applies to the stellar library as a whole, and hence it would not
affect the fitting function residuals. Instead, we aim at constraining
the random errors in the final flux calibration curve and their
effects on the index measurements. 

It is important to note that, in CEN01a, all the stars of a given
observing run were flux calibrated by applying one response curve, the
one that was obtained as an average of all the individual response
curves derived from single observations of spectrophotometric standard
stars (10 -- 20 per observing run). Therefore, using each of the above
individual curves to re-calibrate the stellar spectra, we repeated the
index measurements for all the stars to estimate a random error,
arising from flux calibration uncertainties ($\sigma_{\rm fcal}$), as
the r.m.s. standard deviation of all the individual index
measurements.

The above procedure allowed us to confirm that, as expected (see
Section~\ref{fluxcalx}), and unlike MgI, sTiO is very sensitive to
flux calibration. On average over all the different observing runs, we
obtain that $\sigma_{\rm fcal} \sim 0.012 \times$\,sTiO, what turns
the flux calibration uncertainty into the main source of random error
of the index sTiO (much larger than the joint effect of photon noise
and radial velocity uncertainty). With typical sTiO values in the
range 0.90 -- 0.95, the majority of hot and intermediate $T_{\rm eff}$
stars ($T_{\rm eff} \ga 3900$\,K; $\theta \la 1.30$\,K$^{-1}$) have
$\sigma_{\rm fcal} \sim 0.011$. Colder stars, however, having a much
larger sTiO values, reach errors of up to $\sigma_{\rm fcal} \sim
0.031$. According to the averaged sTiO values of the distinct
categories of stars, different $\sigma_{\rm fcal}$ values for sTiO are
given in Table~\ref{errpredatm}. In addition, a typical value of
$\sigma_{\rm fcal} = 0.007$\,\AA\ for MgI has been derived from all
the library stars.

\subsubsection{Overall random uncertainties}
\label{uncglobal}

The quadratic addition of $\sigma_{\rm typ}$, $\sigma_{\rm par}$ and
$\sigma_{\rm fcal}$ can be interpreted as an overall random
uncertainty, $\sigma_{\rm ALL}$, which is computed and listed for each
group of stars in Table~\ref{errpredatm}. Note that, unlike to what
happens for $\sigma_{\rm typ}$, the new $\sigma_{\rm ALL}$ values
account for most of (or even all) the dispersion of the fits,
$\sigma_{\rm std}$. In other words, the residuals of the fits are
consistent with the expected scatter due to the individual index
errors. This is particularly true for sTiO, with $\sigma_{\rm ALL} =
0.016$ and $\sigma_{\rm std} = 0.017$ for the whole stellar
library. However, it seems that an additional source of error is still
needed to reconcile $\sigma_{\rm ALL}$ and $\sigma_{\rm std}$ for
MgI. This point is addressed in more detail in
Section~\ref{MgFeratio}.

\subsection{[Mg/Fe] abundance ratios}
\label{MgFeratio}

\begin{figure}
\centerline{\hbox{
\psfig{figure=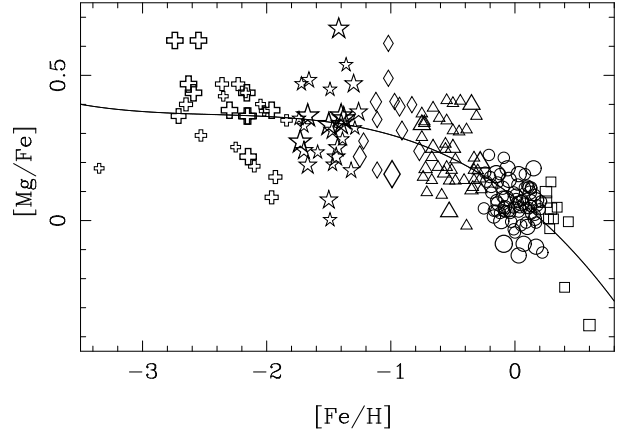,width=8cm,angle=-90}
}}
\caption{[Mg/Fe] abundance ratios vs [Fe/H] for 196 stars of the
library in CEN01a. Different symbols and sizes indicate metallicities
and gravities as in Fig.~\ref{MgIsTiOtheta}a. The solid line is the
least-squares fit to the data given in Eq.~\ref{eq:MgFe-FeH}.}
\label{MgFe}
\end{figure}

\begin{figure}
\centerline{\hbox{
\psfig{figure=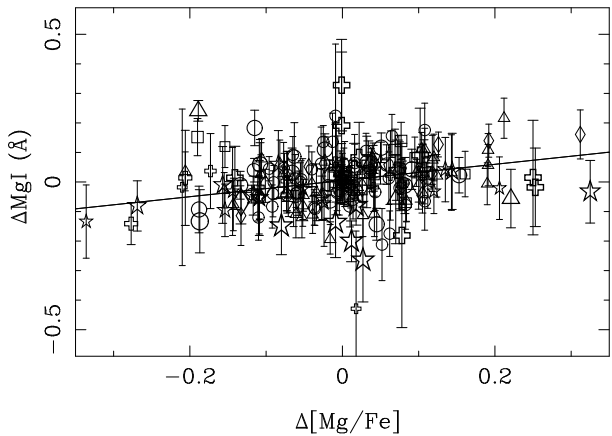,width=8cm,angle=-90}
}} 
\caption{Residuals of the MgI fitting functions ($\Delta$MgI; observed
  -- predicted) vs residuals w.r.t. the intrinsic [Mg/Fe]--[Fe/H]
  relation of the stellar library (Eq.~\ref{eq:MgFe-FeH};
  Figure~\ref{MgFe}). Symbols and sizes as in
  Fig.~\ref{MgIsTiOtheta}a. The solid line represents an
  error-weighted least-squares linear fit to all points within
  3$\sigma$ r.m.s.}
\label{resMgFe}
\end{figure}

Given that MgI is a Mg\,{\sc i} index, in principle one would expect
its metallicity dependence to be better described in terms of the Mg
abundance, [Mg/H], rather than [Fe/H]. In this section we constrain
the importance of different [Mg/Fe] ratios on driving the strength of
the MgI index.

To do this analysis, we have compiled [Mg/Fe] data from the literature
for 196 library stars. Most data were taken from the catalogue of
Borkova \& Marsakov (2005), as they performed a previous compilation
of [Mg/Fe] data in the literature and, then, corrected the different
sources from systematics to end up with an homogeneous [Mg/Fe]
system. In addition, for a few stars which were not available in the
above catalogue, we included the data from Gratton \& Sneden (1987),
Pilachowski, Sneden \& Kraft (1996), and Th\'evenin (1998). The later,
having a large number of stars in common with Borkova \& Marsakov, was
previously transfomed into the Borkova \& Marsakov's system by using
all the stars in common between both catalogues.

Since our stellar library mainly consists of nearby stars, it matches
the well-known [Mg/Fe]-[Fe/H] pattern of the solar neighbourhood
(e.g. Edvardsson et al.~1993), in the sense that [Mg/Fe] decreases
with the increasing [Fe/H]. Figure~\ref{MgFe} illustrates such a trend
for the 196 library stars with available data. Far from being well
represented by just a linear relationship, a least-squares polynomial
fit to all the stars in Fig.~\ref{MgFe} gives

\begin{equation}
\begin{array}{l}
{\rm [Mg/Fe]}_{0}  =  0.08(\pm0.01) - 0.33(\pm0.04)~{\rm [Fe/H]} 
\\ \noalign{\medskip}
- 0.13(\pm0.04)~{\rm [Fe/H]}^2 -0.02(\pm0.01)~{\rm [Fe/H]}^3,
\label{eq:MgFe-FeH}
\end{array}
\end{equation}
where [Mg/Fe]$_{0}$ can be considered as the averaged [Mg/Fe] ratio of
the solar neighbourhood. Note that such a relationship is implicitly
taken into account in the fitting functions through the adopted [Fe/H]
and, therefore, a systematic trend between the MgI fitting functions
residuals and the [Mg/Fe] ratios should not be expected. We have
therefore investigated whether the residuals of the MgI fitting
functions for individual stars ($\Delta {\rm MgI}$; as defined in
Section~\ref{resanderr}) are correlated with their deviations from
Eq.~\ref{eq:MgFe-FeH} ($\Delta {\rm [Mg/Fe]} \equiv {\rm [Mg/Fe]} -
{\rm [Mg/Fe]}_{0}$). Figure~\ref{resMgFe} confirms that there is
indeed a significant relation between $\Delta {\rm MgI}$ and $\Delta
{\rm [Mg/Fe]}$, in the sense that stars with larger Mg abundances (at
fixed [Fe/H]) tend to have positive residuals in the MgI fitting
functions. An error-weighted linear fit to this trend gives
\begin{equation}
\begin{array}{l}
{\Delta {\rm MgI} = 0.005(\pm0.006) + 0.27(\pm0.09)~\Delta {\rm
  [Mg/Fe]}.}
\label{eq:DMgFe-Dres}
\end{array}
\end{equation}
An immediate implication of the above correlation is that it must
account for part of the {\it unexplained} MgI residuals reported in
Section~\ref{uncglobal}. In fact, the variance of the fit in
Eq.~\ref{eq:DMgFe-Dres} is $\sim 10$ percent smaller than
$\sigma^{2}_{\rm std}$ of MgI for the stars in Fig.~\ref{resMgFe}. If
this subsample were representative of the whole library, different
[Mg/Fe] ratios might explain $\sim 0.029$\,\AA\ of the total MgI
$\sigma_{\rm std}$ (0.091\,\AA). As a matter of fact, this number must
be considered as a lower limit since cluster stars do not necessarily
follow the [Mg/Fe] pattern of field stars (see Section~\ref{M67res})
and larger MgI residuals are expected for this subsample of stars.  In
addition, the existence of the correlation given in
Eq.~\ref{eq:DMgFe-Dres} demonstrates that the MgI index is indeed a
good indicator of the Mg abundance. Note that this is not necessarily
true for all metal-line indices, as it was demonstrated in CEN02 that
the Ca\,{\sc ii} triplet indices do not depend on minor changes in
[Ca/Fe].

Unfortunately, [Mg/Fe] ratios are only available for less than
one-third of the library stars. Even though this number is large
enough to perform a reliable analysis as the one presented above, it
is not worthy trying to include a [Mg/Fe] term in our fitting
functions, as not only the small number of stars but also the more
limited atmospheric parameter coverage of the subsample would
dramatically affect the accuracy and reliability of the empirical
calibration. Instead, we prefer to keep the fitting functions in their
present form and stress the idea that, by being constrained to the
chemical enrichment of the solar neighborhood, they are subject to
exhibit systematics with respect to other enrichment scenarios in
which Mg abundances are well different.

\subsubsection{The MgI residual of M67}
\label{M67res}

\begin{figure}
\centerline{\hbox{
\psfig{figure=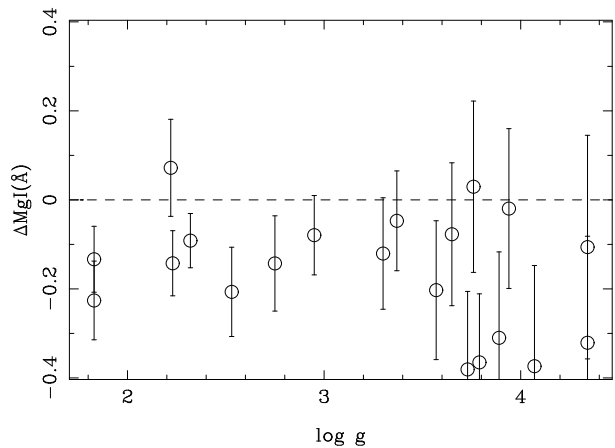,width=8cm,angle=-90}
}}
\caption{MgI residuals ($\Delta$MgI; observed -- predicted) vs $\log
  g$ for M67 stars. A significant offset of $\Delta {\rm MgI} =
  -0.134\pm0.022$\,\AA\ is observed.}
\label{M67off}
\end{figure}

Compared to the predictions of the MgI fitting functions, M67 stars
show a significant mean offset of $\Delta {\rm MgI} =
-0.134\pm0.022$\,\AA\ that cannot be solely explained by typical
random errors. Figure~\ref{M67off} illustrates the derived offsets for
all the 20 stars in M67.

Given that the sensitivity of MgI to $\log g$ is relatively small as
compared to other atmospheric parameters, small uncertainties in the
distance moduli adopted by the original references that derived
surface gravities (see CEN01b) are not expected to be responsible for
the observed discrepancy.  Instead, since MgI increases with
metallicity, the negative MgI residual of M67 could be explained if
the true cluster metallicity were lower than the value adopted in this
work ([Fe/H] $=-0.09$, from Friel \& Janes 1993; see CEN01b). In this
sense, according to our fitting function predictions, a cluster
metallicity of [Fe/H] $\sim -0.32$ would suffice to make the offset
statistically non-significant given the MgI errors of the M67
stars. However, aside the pioneering work by Cohen (1980) in which an
averaged value of [Fe/H] $=-0.39$ was reported for M67, more recent
work derived roughly solar (or slightly below solar) iron abundances
for the cluster stars, like e.g. $-0.1\pm0.1$ (Foy \& Proust 1981),
$-0.09\pm0.07$ (Friel \& Janes 1993), $-0.15 \pm 0.05$ (Friel et
al.~2002), $-0.03\pm0.03$ (Tautvaisiene et al.~2000), $+0.02\pm0.06$
(Gratton 2000). Therefore, assuming that the adopted value is a
reasonable compromise for the [Fe/H] of the cluster, and keeping in
mind that MgI indeed depends on the Mg abundance, the MgI residual
could be a natural consequence of [Mg/Fe] differences between M67
stars and field stars at similar [Fe/H].

As regards to the [Mg/Fe] value of M67, there seems to exist some
discrepancy in the literature. For instance, Tautvaisiene et
al.~(2000) reported [Mg/Fe] $\sim+0.1$ for evolved (giant and clump)
stars in M67, whilst an error-weighted mean of the values presented by
Shetrone \& Sandquist (2000) for a sample of turn-off stars and blue
stragglers in the cluster gives [Mg/Fe] $=-0.14\pm0.10$. 

Making use of Eq.~\ref{eq:MgFe-FeH}, it is easy to convert [Fe/H] into
[Mg/H] abundances to test the above hypothesis.  At [Fe/H] $=-0.09$
(the metallicity adopted for M67), field library stars have an average
value of [Mg/H] $=+0.02$ ([Mg/Fe] $=+0.11$), whereas at [Fe/H] $=
-0.32$ (the value at which the MgI residual would not be significant),
[Mg/H] $=-0.13$ ([Mg/Fe] $=+0.19$). This implies that, if MgI is
indeed driven by Mg rather than Fe, the MgI residual of M67 could be
explained if the Mg content of its stars were offset by $\Delta$[Mg/H]
$=-0.15$ with respect to that of field stars. In other words, keeping
[Fe/H] $= -0.09$ for M67, a value of [Mg/Fe] $\sim -0.04$ for M67 is
required to account for the MgI residuals. This value is in good
agreement with the work by Schiavon, Caldwell \& Rose (2004), which
constrains the [Mg/Fe] of M67 between $-0.1$ and $0.0$ on the basis of
the integrated Mg$b$ in the cluster spectrum. Also, it is consistent
with previous result from Friel \& Janes (1993) that the metallicity
derived for M67 from Mg$b$ is significantly lower than the one
inferred from Fe indices. We therefore conclude that the MgI residual
of M67 is probably driven by the existence of a Mg underabundance of
[Mg/Fe] $\sim -0.04$, in contrast with the averaged value [Mg/Fe]
$\sim +0.11$ for field library stars of equal [Fe/H].

\section{A comparison with theoretical work}
\label{theoretical}

\begin{figure*}
\centerline{\hbox{
\psfig{figure=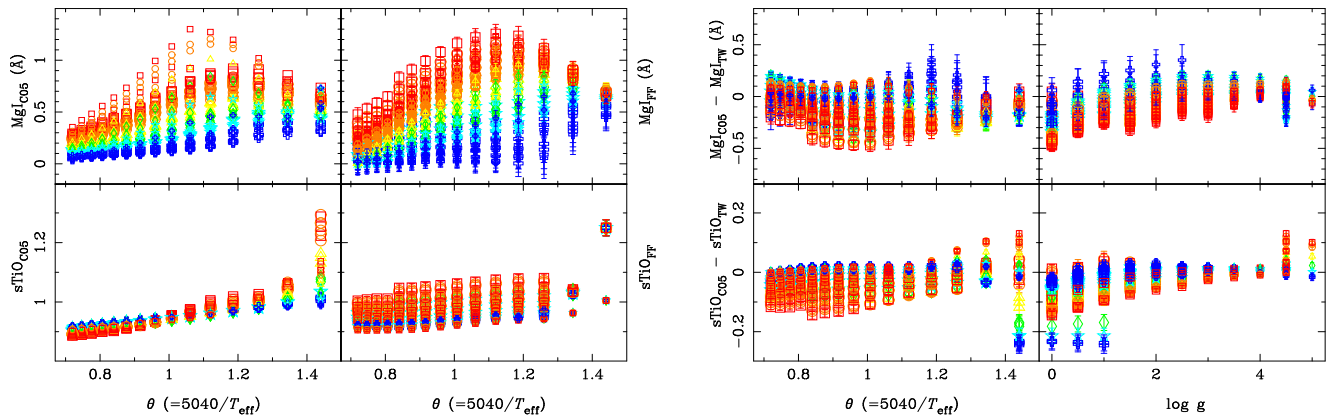,width=17.5cm}
}}
\caption{{\it Left four panels:} A comparison between the MgI and sTiO
  indices measured for the synthetic stars of Coelho et al.~(2005;
  C05) and those predicted by the fitting functions (FF) of this paper
  at exactly the same atmospheric parameters of the synthetic
  stars. {\it Right four panels:} Star-by-star differences between the
  C05 and FF indices versus $\theta$ and $\log g$. For each index,
  vertical scales have been set to the same relative ranges of left
  panels to allow direct comparisons. Symbol types and relative sizes
  (indicating, respectively, metallicity and gravity ranges) are given
  as in Fig.~\ref{MgIsTiOtheta}a. See more details in the text.}
\label{CoelhovsFFs}
\end{figure*}

We here provide a qualitative comparison between the new fitting
functions presented in this paper and the MgI and sTiO predictions
derived from theoretical work based on model atmospheres, in
particular from the high-resolution synthetic stellar spectral library
of Coelho et al.~(2005; hereafter C05). This theoretical library
ranges from the near-ultraviolet to the near-infrared, spaning the
atmospheric parameter range of $3500 \leq T_{\rm eff} \leq 7000$\,K,
$0.0 \leq \log g \leq 5.0$\,dex, and $-2.5 \leq $[Fe/H]$ \leq
+0.5$\,dex, at two different $\alpha$-element abundance ratios,
[$\alpha$/Fe] = 0.0 and $+0.4$.

Prior to measuring the MgI and sTiO indices over the C05 stellar
spectra, we smoothed and rebinned the C05 stellar spectra to match the
spectral resolution and linear dispersion of the stars in CEN01a
(FWHM$ = 1.5$~\AA; 0.85\,\AA/pix). Also, for the sake of carrying out
a meaningful comparison with our predictions, we removed from the
current analysis all those synthetic stars that, because of their
atmospheric parameters, are not expected to be required by SSP
codes. This includes, for instance, late spectral types with
intermediate surface gravity values (e.g. from $T_{\rm eff} = 4500$\,K
and $\log g = 4$\,dex, down to $T_{\rm eff} = 3500$\,K and $1.5 \leq
\log g \leq 4$\,dex), and hot dwarfs with very high surface gravity
values ($T_{\rm eff} \geq 5000$\,K and $\log g = 5$). This way we also
guaranteed that the overall atmospheric parameter space of the
synthetic stars is not very different from that covered by the real
stars in CEN01a.

To avoid systematics between the MgI of both samples arising from the
existence of different Mg abundances at fixed [Fe/H], we took into
account the [Mg/Fe]-[Fe/H] abundance pattern of the stars in CEN01a
(see Section~\ref{MgFeratio}) and corrected the C05 indices for this
effect. To do this, for each [Fe/H] value in C05 we determined its
corresponding [Mg/Fe] according to Eq.~\ref{eq:MgFe-FeH}. Hence, the
final MgI and sTiO indices for the synthetic C05 stellar spectra are
the result of interpolating linearly the indices measured at
[$\alpha$/Fe] = 0.0 and 0.4 according to the above [Mg/Fe] value.

With all the considerations given above, the four left panels in
Figure~\ref{CoelhovsFFs} illustrate a comparison between the MgI and
sTiO indices measured on the C05 synthetic stars and those derived
from our MgI and sTiO fitting functions for exactly the same
atmospheric parameters of the synthetic stars. Although there exists a
reasonable qualitative agreement in the first order behaviours of both
datasets, absolute numbers reveal significant differences between real
and synthetic indices. This is more clearly presented in the four
right panels of Fig.~\ref{CoelhovsFFs}, where the star-by-star
absolute differences are plotted versus $\theta$ and $\log g$. The
r.m.s.~standard deviation of the index differences are $\sigma_{{\rm
std}} = 0.142$\,\AA\ and 0.042 for MgI and sTiO respectively, much
larger than the typical residuals in Table~\ref{sigmas} and
Figs.~\ref{residualsMgI} and~\ref{residualssTiO}. Some of these
differences may be due to the limitation of our fitting functions to
reproduce the poorly populated regions of the parameter space
(e.g.~the MgI of metal-poor stars at $\theta \sim 1.2$; see discussion
in Section~\ref{mgiffs}). However, most cases can not be justified in
this way (e.g.~synthetic MgI and sTiO indices do not to reproduce
satisfactorily the gravity dependence exhibited by real stars,
particularly for intermediate-to-high metallicity giants) and the
intrinsic limitations of theoretical model atmospheres hence appear as
potential sources for the observed differences. It is not however the
scope of this section to provide a detailed analysis of the origins
for the observed differences but to illustrate the reader with a
comparative analysis between theoretical and empirical work. A similar
comparison for the Ca\,{\sc ii} triplet lines can be found in Vazdekis
et al.~(2003).

\section{Summary and conclusions}
\label{conclusions}

Based on the near-IR stellar library of CEN01a,b we have defined new
line-strength indices for the MgI line at 8807~\AA\ and the TiO bands
around the Ca\,{\sc ii} triplet region.  These indices, called MgI and
sTiO respectively, are thought to be used as stellar population
diagnostics. For this reason, we have characterized their
sensitivities to different signal-to-noise ratios, distinct spectral
resolutions, flux calibration and reddening correction systematics,
and the presence of sky line residuals typical of this spectral
range. Also, we give some recipes for those readers interested in
transforming their old MgI and TiO index data into our new system of
indices.

After measuring MgI and sTiO for all the library stars at the spectral
resolution of CEN01a, we have calibrated their dependences on the
stellar atmospheric parameters ($T_{\rm eff}$, $\log g$ and [Fe/H]) by
means of empirical fitting functions. The reliability and accuracy of
the new fitting function predictions have been discussed by performing
a thorough analysis of the different error sources driving the fitting
function residuals.  For sTiO, these residuals are overall compatible
with the ones expected from the uncertainties in the input atmospheric
parameters and from the random errors in the index measurements
(accounting for photon noise, radial velocity uncertainties, and flux
calibration errors). For MgI, however, an additional contribution to
the fitting function residuals arises from the existence of distinct
[Mg/Fe] ratios among the library stars. As a consequence of this
analysis, we have detected a statistically significant offset in the
MgI values of M67 stars. This is consistent with M67 having a [Mg/Fe]
underabundance of $\sim -0.04$, in contrast with the typical
[Mg/Fe]$\sim +0.11$ of solar-neighbourhood stars with the same [Fe/H].

A full database with the index measurements for each library star,
their random errors, the fitting function residuals, and the compiled
[Mg/Fe] values is given in Table~\ref{macrotab}.  This database,
together with the {\tt FORTRAN} routine for the evaluation of the
fitting functions, are also available at \\ {\tt
http://www.ucm.es/info/Astrof/ellipt/MgIsTiO.html}\\ In a forthcoming
paper (Vazdekis et al. in preparation), SSP model predictions for MgI
and sTiO will be presented on the basis of such fitting functions,
either at the above website and at A.~Vazdekis models
website\footnote{http://www.iac.es/galeria/vazdekis/vazdekis\_models.html}.

To conclude, a brief summary of the main characteristics of the
indices, their behaviours, and their potential capabilities for
stellar population studies is given below:

\begin{itemize}

\item MgI has been specifically designed to be a very sensitive
indicator of Mg abundance. For this reason, because of its relatively
narrow line bandpass, it turns out to be dependent on the overall
spectral resolution (and, hence, on the velocity dispersion of
galaxies). Users are therefore encouraged to put their data into a
homogeneous system of spectral resolution (either using the
sigma-dependent polynomials provided in Table~\ref{polcoefdisp}, or
broadening their spectra up to a common overall spectral resolution)
before any meaningful comparison of the MgI index among different
types of objects.

MgI exhibits a complex dependence on the three main stellar
atmospheric parameters. For hot and cold stars, $T_{\rm eff}$ and
$\log g$ are the main driving parameters, whereas, in the
mid-temperature regime ($4000 \leq T_{\rm eff} \leq 7000$\,K), all
three parameters play a significant role: MgI increases with the
increasing metallicity and the decreasing temperature, with a mild
dependence on the luminosity class that makes dwarfs and supergiant
stars to exhibit slightly larger MgI values than giant stars.

For stellar populations studies, as it will be shown in a forthcoming
paper (Vazdekis et al.~in prep.), the integrated MgI turns out to be
an excellent indicator of the Mg abundance.

\item The sTiO index has been defined to measure the slope of the
pseudo-continuum of the Ca\,{\sc ii} triplet region, which is known to
change dramatically for M-type stars due to the appearence of strong
TiO molecular bands. Because of its definition, the sTiO index is
strikingly robust against changes in spectral resolution ---and
velocity dispersions--- and low S/N ratios, which makes it
particularly suited for the analysis of galaxies at high redshifts and
extragalactic globular clusters. It however requires the spectra to
have a proper relative flux calibration.

As regards to the behaviour of sTiO in stars, it is worth noting the
steep increase with the decreasing temperature for $T_{\rm eff} \leq
3600$\,K. In turn, at such low temperatures, giant stars exhibit much
stronger sTiO values than dwarfs, what makes sTiO to be a powerful
dwarf-to-giant discriminator for M-type stars. In addition, the
dependence of sTiO on the stellar metallicity is almost negligible.

For the integrated spectra of quiescent galaxies, however, sTiO is
found to reflect the overall metallicity of the stellar population, as
the red giant branch gets cooler with the increasing metallicity and
the relative contribution of M-type stars increases (see Vazdekis et
al.~2003). 

\end{itemize}

In Cenarro et al.~(2003), both MgI and sTiO were measured for the
first time over a sample of 35 elliptical galaxies, illustrating the
usefulness of the MgI index to reproduce the Mg-$\sigma$ relation of
elliptical galaxies at the near-IR, and the capabilities of sTiO as
metallicity indicator.

\section*{ACKNOWLEDGMENTS}

We acknowledge the anonymous referee for very useful comments.
A.J.C. is a {\it Juan de la Cierva} Fellow of the Spanish Ministry of
Science and Innovation. This work has been funded by the Spanish
Ministry of Science and Innovation through grants AYA2007-67752-C03-01
and AYA2006-15698-C02-02.


\include{apen_tabla_rev}

\end{document}

%% file: apen_tabla_rev.tex
\begin{table*}
\caption{\small MgI and sTiO index measurements ($\cal I$), typical
  observational index error ($\sigma_{\rm typ}\cal I$, accounting for
  photon noise and typical uncertainties in radial velocity), fitting
  function residuals ($\Delta\cal I$), and [Mg/Fe] abundances (when
  available) for all the stars in CEN01a. Sources for [Mg/Fe] values
  are: (1) the catalogue of spectroscopic abundances in stars (Borkova
  \& Marsakov 2005); (2) the catalogue of chemical abundances in
  late-type stars (Thevenin 1998) corrected to the Borkova \& Marsakov
  (2005) system; (3) Pilachowski, Sneden \& Kraft (1996); and (4)
  Gratton \& Sneden (1987).  The first five columns indicates the star
  number within the library, the name, the coordinates (RA and Dec in
  J2000, except M71 stars wich are B1950), and the spectral type
  and luminosity class. Additional information for the library stars
  (e.g.~atmospheric parameters, [Ca/Fe] abundances, CaT, PaT, and
  CaT$^*$ index values, their errors and fitting function residuals,
  $B$ and $V$ apparent magnitudes, signal-to-noise ratios per
  angstrom, and other names) can be found at CEN01a,b and at {\tt
    http://www.ucm.es/info/Astrof/ellipt/MgIsTiO.html}.}
\label{macrotab}%
\centering{\small %
\begin{tabular}{llcclr@{}r@{}rr@{}r@{}rr}
\hline\hline
Num& Name                        &RA&Dec& SpT         & \multicolumn{3}{c}{MgI(\AA)} & \multicolumn{3}{c}{sTiO}& [Mg/Fe] \\                              
   &               &(h:m:s)  &(d:m:s)   &             & $\cal I$\ \ \ &$\sigma_{\rm typ}\cal I$\ &$\Delta\cal I$\ \ \ & $\cal I$\ \ \ &$\sigma_{\rm typ}\cal I$\ &$\Delta\cal I$\ \ \ \\                              
\hline
001&Coma A   3                         &12:19:16.7&$+$25:26:11&G9 V         &    0.417&\ \ 0.097&\ $-$0.430& 0.900&\ \ 0.008&\ $-$0.025&               \\
002&Coma A  13                         &12:24:06.2&$+$26:07:44&K0 V         &    0.388&\ \ 0.135&\ $-$0.318& 0.865&\ \ 0.010&\ $-$0.053&               \\
003&Coma A  14                         &12:24:17.6&$+$24:19:32&G4 V         &    0.396&\ \ 0.080&\ $-$0.297& 0.920&\ \ 0.007&\ $-$0.002&               \\
004&Coma A  21                         &12:30:14.8&$+$25:01:44&G7 V         &    0.570&\ \ 0.099&\ $-$0.185& 0.919&\ \ 0.008&\ $-$0.005&               \\ \medskip
005&Coma T  65                         &12:19:28.3&$+$24:17:03&G0 V         &    0.285&\ \ 0.079&\ $-$0.143& 0.896&\ \ 0.006&\ $-$0.017&               \\
006&Coma T  68                         &12:19:50.6&$+$28:27:51&A6 IV-V      & $-$0.064&\ \ 0.066&\ $-$0.008& 0.920&\ \ 0.005&\ $-$0.016&               \\
\hline
\end{tabular}
}                                                              
\end{table*}

\begin{table*}
\addtocounter{table}{-1}
\caption{\small Continuation}
\label{macrotab}%
\centering{\small %
\begin{tabular}{llcclr@{}r@{}rr@{}r@{}rr}
\hline\hline
Num& Name                        &RA&Dec& SpT         & \multicolumn{3}{c}{MgI(\AA)} & \multicolumn{3}{c}{sTiO}& [Mg/Fe] \\                              
   &               &(h:m:s)  &(d:m:s)   &             & $\cal I$\ \ \ &$\sigma_{\rm typ}\cal I$\ &$\Delta\cal I$\ \ \ & $\cal I$\ \ \ &$\sigma_{\rm typ}\cal I$\ &$\Delta\cal I$\ \ \ \\                              
\hline
051&M5  IV-19                          &15:18:33.1&$+$02:01:18&GB           &    0.266&\ \ 0.232&\ $-$0.417& 0.920&\ \ 0.019&\ $-$0.070&               \\
052&M5  IV-59                          &15:18:44.0&$+$02:04:36&GB           &    0.573&\ \ 0.197&\ $-$0.054& 0.946&\ \ 0.018&\ $-$0.043&               \\
053&M5  IV-86                          &15:18:43.0&$+$02:01:49&HB           &    0.228&\ \ 0.202&\    0.022& 0.924&\ \ 0.017&\ $-$0.004&               \\
054&M5  IV-87                          &15:18:44.2&$+$02:01:52&HB           &    0.146&\ \ 0.127&\    0.009& 0.936&\ \ 0.011&\    0.012&               \\ \medskip
055&M67 F 084                          &08:51:12.7&$+$11:52:42&HB           &    0.620&\ \ 0.061&\ $-$0.092& 0.926&\ \ 0.005&\ $-$0.032&               \\
056&M67 F 094                          &08:51:15.4&$+$11:50:14&SGB          & $-$0.078&\ \ 0.226&\ $-$0.374& 0.873&\ \ 0.017&\ $-$0.042&               \\
\hline
\end{tabular}
}                                                              
\end{table*}

\begin{table*}
\addtocounter{table}{-1}
\caption{\small Continuation}
\label{macrotab}%
\centering{\small %
\begin{tabular}{llcclr@{}r@{}rr@{}r@{}rr}
\hline\hline
Num& Name                        &RA&Dec& SpT         & \multicolumn{3}{c}{MgI(\AA)} & \multicolumn{3}{c}{sTiO}& [Mg/Fe] \\                              
   &               &(h:m:s)  &(d:m:s)   &             & $\cal I$\ \ \ &$\sigma_{\rm typ}\cal I$\ &$\Delta\cal I$\ \ \ & $\cal I$\ \ \ &$\sigma_{\rm typ}\cal I$\ &$\Delta\cal I$\ \ \ \\                              
\hline
106&M71 Z                              &          &$ $        &             &    0.234&\ \ 0.230&\         & 0.931&\ \ 0.020&\         &               \\
107&M71 KC 22                          &          &$ $        &             &    0.534&\ \ 0.174&\         & 0.927&\ \ 0.015&\         &               \\
108&M71 KC 147                         &19:51:30  &$+$18:37:13&             &    0.591&\ \ 0.165&\    0.133& 0.942&\ \ 0.015&\ $-$0.002&               \\
109&M71 KC 169                         &19:51:40  &$+$18:37:03&             &    0.503&\ \ 0.158&\    0.062& 0.933&\ \ 0.014&\ $-$0.010&               \\ \medskip
110&M71 KC 263                         &19:51:31  &$+$18:39:13&             &    0.294&\ \ 0.193&\ $-$0.168& 0.898&\ \ 0.016&\ $-$0.046&               \\
111&M92 I-10                           &17:17:36.4&$+$43:09:32&HB           & $-$0.579&\ \ 0.274&\ $-$0.384& 0.971&\ \ 0.022&\    0.000&               \\
\hline
\end{tabular}
}                                                              
\end{table*}

\begin{table*}
\addtocounter{table}{-1}
\caption{\small Continuation}
\label{macrotab}%
\centering{\small %
\begin{tabular}{llcclr@{}r@{}rr@{}r@{}rr}
\hline\hline
Num& Name                        &RA&Dec& SpT         & \multicolumn{3}{c}{MgI(\AA)} & \multicolumn{3}{c}{sTiO}& [Mg/Fe] \\                              
   &               &(h:m:s)  &(d:m:s)   &             & $\cal I$\ \ \ &$\sigma_{\rm typ}\cal I$\ &$\Delta\cal I$\ \ \ & $\cal I$\ \ \ &$\sigma_{\rm typ}\cal I$\ &$\Delta\cal I$\ \ \ \\                              
\hline
161&HD\,1918                           &00:23:43.2&$+$45:05:21&G9 III       &    0.535&\ \ 0.052&\    0.000& 0.954&\ \ 0.005&\ $-$0.001&               \\
162&HD\,2665                           &00:30:45.4&$+$57:03:54&G5 IIIwe     & $-$0.009&\ \ 0.044&\ $-$0.141& 0.952&\ \ 0.004&\    0.021&   +0.08 $^{3}$\\
163&HD\,2857                           &00:31:53.8&$-$05:15:43&A2 (HB)      & $-$0.004&\ \ 0.093&\    0.105& 0.959&\ \ 0.008&\    0.016&               \\
164&HD\,3443                           &00:37:20.7&$-$24:46:02&K1 V + ...   &    0.600&\ \ 0.080&\ $-$0.065& 0.923&\ \ 0.007&\    0.005&               \\ \medskip
165&HD\,3546                           &00:38:33.3&$+$29:18:42&G5 III       &    0.473&\ \ 0.012&\    0.010& 0.936&\ \ 0.001&\ $-$0.006&   +0.25 $^{2}$\\
166&HD\,3567                           &00:38:31.9&$-$08:18:33&F5 V         &    0.046&\ \ 0.061&\ $-$0.096& 0.913&\ \ 0.005&\ $-$0.008&   +0.172$^{1}$\\
\hline
\end{tabular}
}                                                              
\end{table*}

\begin{table*}
\addtocounter{table}{-1}
\caption{\small Continuation}
\label{macrotab}%
\centering{\small %
\begin{tabular}{llcclr@{}r@{}rr@{}r@{}rr}
\hline\hline
Num& Name                        &RA&Dec& SpT         & \multicolumn{3}{c}{MgI(\AA)} & \multicolumn{3}{c}{sTiO}& [Mg/Fe] \\                              
   &               &(h:m:s)  &(d:m:s)   &             & $\cal I$\ \ \ &$\sigma_{\rm typ}\cal I$\ &$\Delta\cal I$\ \ \ & $\cal I$\ \ \ &$\sigma_{\rm typ}\cal I$\ &$\Delta\cal I$\ \ \ \\                              
\hline
216&HD\,19476                          &03:09:29.8&$+$44:51:27&K0 III       &    0.671&\ \ 0.031&\ $-$0.072& 0.951&\ \ 0.003&\    0.004&               \\
217&HD\,20041                          &03:15:48.0&$+$57:08:26&A0 Ia        & $-$0.093&\ \ 0.032&\         & 0.963&\ \ 0.003&\         &               \\
218&HD\,20630                          &03:19:21.7&$+$03:22:13&G5 V         &    0.504&\ \ 0.041&\ $-$0.086& 0.898&\ \ 0.003&\ $-$0.016&   +0.031$^{1}$\\
219&HD\,20893                          &03:22:45.2&$+$20:44:31&K3 III       &    0.933&\ \ 0.034&\    0.050& 0.981&\ \ 0.003&\    0.006&               \\ \medskip
220&HD\,22049                          &03:32:55.8&$-$09:27:30&K2 V         &    0.738&\ \ 0.053&\ $-$0.040& 0.931&\ \ 0.005&\    0.007&   +0.034$^{1}$\\
221&HD\,22484                          &03:36:52.4&$+$00:24:06&F8 V         &    0.345&\ \ 0.033&\ $-$0.038& 0.911&\ \ 0.003&\ $-$0.005&   +0.046$^{1}$\\
\hline
\end{tabular}
}                                                              
\end{table*}

\begin{table*}
\addtocounter{table}{-1}
\caption{\small Continuation}
\label{macrotab}%
\centering{\small %
\begin{tabular}{llcclr@{}r@{}rr@{}r@{}rr}
\hline\hline
Num& Name                        &RA&Dec& SpT         & \multicolumn{3}{c}{MgI(\AA)} & \multicolumn{3}{c}{sTiO}& [Mg/Fe] \\                              
   &               &(h:m:s)  &(d:m:s)   &             & $\cal I$\ \ \ &$\sigma_{\rm typ}\cal I$\ &$\Delta\cal I$\ \ \ & $\cal I$\ \ \ &$\sigma_{\rm typ}\cal I$\ &$\Delta\cal I$\ \ \ \\                              
\hline
271&HD\,49161                          &06:47:19.8&$+$08:02:14&K4 III       &    1.006&\ \ 0.033&\    0.071& 0.982&\ \ 0.003&\ $-$0.011&               \\
272&HD\,49293                          &06:47:51.6&$+$02:24:44&K0 III       &    0.734&\ \ 0.041&\ $-$0.043& 0.973&\ \ 0.004&\    0.009&               \\
273&HD\,50778                          &06:54:11.4&$-$12:02:19&K4 III       &    0.865&\ \ 0.032&\    0.012& 0.975&\ \ 0.003&\ $-$0.007&               \\
274&HD\,51440                          &06:59:02.8&$+$38:03:08&K2 III       &    0.722&\ \ 0.012&\    0.000& 0.959&\ \ 0.001&\ $-$0.005&               \\ \medskip
275&HD\,52005                          &07:00:15.8&$+$16:04:44&K4 Iab       &    1.065&\ \ 0.034&\    0.033& 1.008&\ \ 0.003&\ $-$0.038&               \\
276&HD\,52973                          &07:04:06.5&$+$20:34:13&G0 Ib var    &    0.096&\ \ 0.063&\ $-$0.478& 0.934&\ \ 0.005&\ $-$0.026&               \\
\hline
\end{tabular}
}                                                              
\end{table*}

\begin{table*}
\addtocounter{table}{-1}
\caption{\small Continuation}
\label{macrotab}%
\centering{\small %
\begin{tabular}{llcclr@{}r@{}rr@{}r@{}rr}
\hline\hline
Num& Name                        &RA&Dec& SpT         & \multicolumn{3}{c}{MgI(\AA)} & \multicolumn{3}{c}{sTiO}& [Mg/Fe] \\                              
   &               &(h:m:s)  &(d:m:s)   &             & $\cal I$\ \ \ &$\sigma_{\rm typ}\cal I$\ &$\Delta\cal I$\ \ \ & $\cal I$\ \ \ &$\sigma_{\rm typ}\cal I$\ &$\Delta\cal I$\ \ \ \\                              
\hline
326&HD\,82328                          &09:32:51.4&$+$51:40:38&F6 IV        &    0.276&\ \ 0.026&\    0.040& 0.918&\ \ 0.002&\    0.001&   +0.014$^{1}$\\
327&HD\,82885                          &09:35:39.5&$+$35:48:36&G8 IV--V     &    0.744&\ \ 0.041&\    0.075& 0.915&\ \ 0.004&\    0.003&               \\
328&HD\,83618                          &09:39:51.4&$-$01:08:34&K3 III       &    0.775&\ \ 0.059&\ $-$0.089& 0.968&\ \ 0.006&\ $-$0.013&               \\
329&HD\,84441                          &09:45:51.1&$+$23:46:27&G1 II        &    0.432&\ \ 0.040&\ $-$0.119& 0.935&\ \ 0.003&\ $-$0.019&               \\ \medskip
330&HD\,84737                          &09:48:35.4&$+$46:01:16&G2V          &    0.464&\ \ 0.063&\    0.013& 0.920&\ \ 0.005&\    0.005&   +0.113$^{1}$\\
331&HD\,84937                          &09:48:56.1&$+$13:44:39&F5 VI        &    0.020&\ \ 0.041&\ $-$0.007& 0.916&\ \ 0.003&\ $-$0.008&   +0.440$^{1}$\\
\hline
\end{tabular}
}                                                              
\end{table*}

\begin{table*}
\addtocounter{table}{-1}
\caption{\small Continuation}
\label{macrotab}%
\centering{\small %
\begin{tabular}{llcclr@{}r@{}rr@{}r@{}rr}
\hline\hline
Num& Name                        &RA&Dec& SpT         & \multicolumn{3}{c}{MgI(\AA)} & \multicolumn{3}{c}{sTiO}& [Mg/Fe] \\                              
   &               &(h:m:s)  &(d:m:s)   &             & $\cal I$\ \ \ &$\sigma_{\rm typ}\cal I$\ &$\Delta\cal I$\ \ \ & $\cal I$\ \ \ &$\sigma_{\rm typ}\cal I$\ &$\Delta\cal I$\ \ \ \\                              
\hline
381&HD\,113226                         &13:02:10.6&$+$10:57:33&G8 IIIvar    &    0.734&\ \ 0.049&\    0.059& 0.922&\ \ 0.004&\ $-$0.023&   +0.06 $^{2}$\\
382&HD\,113285                         &13:02:38.0&$+$05:11:08&M8 III       &    0.146&\ \ 0.021&\         & 2.562&\ \ 0.007&\         &               \\
383&HD\,113848                         &13:06:21.2&$+$21:09:12&F4 V         &    0.214&\ \ 0.045&\    0.056& 0.915&\ \ 0.004&\ $-$0.001&               \\
384&HD\,114710                         &13:11:52.4&$+$27:52:41&G0 V         &    0.466&\ \ 0.043&\    0.009& 0.929&\ \ 0.004&\    0.019& $-$0.007$^{1}$\\ \medskip
385&HD\,114762                         &13:12:19.7&$+$17:31:02&F9 V         &    0.169&\ \ 0.051&\ $-$0.113& 0.905&\ \ 0.004&\ $-$0.014&   +0.287$^{1}$\\
386&HD\,114946                         &13:14:10.9&$-$19:55:51&G6 V         &    0.353&\ \ 0.046&\ $-$0.328& 0.941&\ \ 0.004&\    0.003&               \\
\hline
\end{tabular}
}                                                              
\end{table*}

\begin{table*}
\addtocounter{table}{-1}
\caption{\small Continuation}
\label{macrotab}%
\centering{\small %
\begin{tabular}{llcclr@{}r@{}rr@{}r@{}rr}
\hline\hline
Num& Name                        &RA&Dec& SpT         & \multicolumn{3}{c}{MgI(\AA)} & \multicolumn{3}{c}{sTiO}& [Mg/Fe] \\                              
   &               &(h:m:s)  &(d:m:s)   &             & $\cal I$\ \ \ &$\sigma_{\rm typ}\cal I$\ &$\Delta\cal I$\ \ \ & $\cal I$\ \ \ &$\sigma_{\rm typ}\cal I$\ &$\Delta\cal I$\ \ \ \\                              
\hline
436&HD\,135722                         &15:15:30.2&$+$33:18:53&G8 III       &    0.562&\ \ 0.034&\    0.010& 0.944&\ \ 0.003&\ $-$0.004&               \\
437&HD\,136028                         &15:18:26.1&$-$00:27:40&K5 III       &    0.828&\ \ 0.079&\ $-$0.066& 0.974&\ \ 0.008&\ $-$0.007&               \\
438&HD\,136202                         &15:19:18.8&$+$01:45:55&F8 III-IV    &    0.283&\ \ 0.052&\ $-$0.038& 0.909&\ \ 0.004&\ $-$0.008&   +0.046$^{1}$\\
439&HD\,136479                         &15:21:07.6&$-$05:49:29&K1 III       &    0.813&\ \ 0.034&\    0.016& 0.951&\ \ 0.003&\ $-$0.005&               \\ \medskip
440&HD\,136726                         &15:17:05.9&$+$71:49:26&K4 III       &    0.979&\ \ 0.056&\    0.070& 0.982&\ \ 0.005&\    0.001&               \\
441&HD\,137391                         &15:24:29.4&$+$37:22:38&F0 V         &    0.103&\ \ 0.040&\ $-$0.032& 0.929&\ \ 0.003&\    0.015&               \\
\hline
\end{tabular}
}                                                              
\end{table*}

\begin{table*}
\addtocounter{table}{-1}
\caption{\small Continuation}
\label{macrotab}%
\centering{\small %
\begin{tabular}{llcclr@{}r@{}rr@{}r@{}rr}
\hline\hline
Num& Name                        &RA&Dec& SpT         & \multicolumn{3}{c}{MgI(\AA)} & \multicolumn{3}{c}{sTiO}& [Mg/Fe] \\                              
   &               &(h:m:s)  &(d:m:s)   &             & $\cal I$\ \ \ &$\sigma_{\rm typ}\cal I$\ &$\Delta\cal I$\ \ \ & $\cal I$\ \ \ &$\sigma_{\rm typ}\cal I$\ &$\Delta\cal I$\ \ \ \\                              
\hline
491&HD\,159307                         &17:34:08.4&$-$03:03:20&F8           &    0.189&\ \ 0.045&\    0.031& 0.920&\ \ 0.004&\    0.002&   +0.178$^{1}$\\
492&HD\,159332                         &17:33:22.8&$+$19:15:24&F6 V         &    0.285&\ \ 0.036&\    0.022& 0.923&\ \ 0.003&\    0.006&   +0.119$^{1}$\\
493&HD\,159561                         &17:34:56.1&$+$12:33:36&A5 III       & $-$0.091&\ \ 0.035&\ $-$0.028& 0.947&\ \ 0.003&\    0.008&               \\
494&HD\,160365                         &17:38:57.9&$+$13:19:45&F6 III       &    0.259&\ \ 0.037&\         & 0.926&\ \ 0.003&\         &               \\ \medskip
495&HD\,160693                         &17:39:36.9&$+$37:11:02&G0 V         &    0.402&\ \ 0.058&\    0.075& 0.913&\ \ 0.005&\ $-$0.006&   +0.353$^{1}$\\
496&HD\,161797                         &17:46:27.5&$+$27:43:14&G5 IV        &    0.556&\ \ 0.048&\ $-$0.066& 0.927&\ \ 0.004&\    0.004&   +0.006$^{1}$\\
\hline
\end{tabular}
}                                                              
\end{table*}

\begin{table*}
\addtocounter{table}{-1}
\caption{\small Continuation}
\label{macrotab}%
\centering{\small %
\begin{tabular}{llcclr@{}r@{}rr@{}r@{}rr}
\hline\hline
Num& Name                        &RA&Dec& SpT         & \multicolumn{3}{c}{MgI(\AA)} & \multicolumn{3}{c}{sTiO}& [Mg/Fe] \\                              
   &               &(h:m:s)  &(d:m:s)   &             & $\cal I$\ \ \ &$\sigma_{\rm typ}\cal I$\ &$\Delta\cal I$\ \ \ & $\cal I$\ \ \ &$\sigma_{\rm typ}\cal I$\ &$\Delta\cal I$\ \ \ \\                              
\hline
546&HD\,175751                         &18:57:03.7&$-$05:50:47&K2 III       &    0.777&\ \ 0.048&\    0.028& 0.966&\ \ 0.004&\    0.009&               \\
547&HD\,175865                         &18:55:20.1&$+$43:56:46&M5 III       &    0.735&\ \ 0.055&\ $-$0.021& 1.655&\ \ 0.010&\    0.099&               \\
548&HD\,176301                         &18:58:45.1&$+$19:47:40&B7 III-IV    & $-$0.220&\ \ 0.037&\         & 0.962&\ \ 0.003&\         &               \\
549&HD\,176411                         &18:59:37.4&$+$15:04:06&K2 III       &    1.015&\ \ 0.033&\    0.246& 0.969&\ \ 0.003&\    0.013&               \\ \medskip
550&HD\,178717                         &19:09:22.0&$+$10:14:28&K3.5 III     &    0.732&\ \ 0.039&\ $-$0.138& 1.048&\ \ 0.004&\    0.048&               \\
551&HD\,180711                         &19:12:33.3&$+$67:39:42&G9 III       &    0.629&\ \ 0.057&\ $-$0.047& 0.958&\ \ 0.005&\    0.008&               \\
\hline
\end{tabular}
}                                                              
\end{table*}

\begin{table*}
\addtocounter{table}{-1}
\caption{\small Continuation}
\label{macrotab}%
\centering{\small %
\begin{tabular}{llcclr@{}r@{}rr@{}r@{}rr}
\hline\hline
Num& Name                        &RA&Dec& SpT         & \multicolumn{3}{c}{MgI(\AA)} & \multicolumn{3}{c}{sTiO}& [Mg/Fe] \\                              
   &               &(h:m:s)  &(d:m:s)   &             & $\cal I$\ \ \ &$\sigma_{\rm typ}\cal I$\ &$\Delta\cal I$\ \ \ & $\cal I$\ \ \ &$\sigma_{\rm typ}\cal I$\ &$\Delta\cal I$\ \ \ \\                              
\hline
601&HD\,199191                         &20:53:18.3&$+$54:31:05&K0 III       &    0.623&\ \ 0.055&\    0.049& 0.941&\ \ 0.005&\ $-$0.010&               \\
602&HD\,199478                         &20:55:49.8&$+$47:25:04&B8 Iae       & $-$0.056&\ \ 0.026&\         & 0.928&\ \ 0.002&\         &               \\
603&HD\,199580                         &20:56:48.0&$+$42:53:44&K0 III-IV    &    0.639&\ \ 0.031&\    0.001& 0.948&\ \ 0.003&\    0.013&               \\
604&HD\,199960                         &21:00:33.8&$-$04:43:49&G1 V         &    0.569&\ \ 0.042&\    0.040& 0.904&\ \ 0.004&\ $-$0.010&   +0.052$^{1}$\\ \medskip
605&HD\,200580                         &21:04:07.4&$+$02:59:40&F9 V         &    0.422&\ \ 0.047&\    0.079& 0.927&\ \ 0.004&\    0.008&   +0.139$^{1}$\\
606&HD\,200779                         &21:05:19.7&$+$07:04:09&K6 V         &    1.093&\ \ 0.026&\         & 0.946&\ \ 0.002&\         &               \\
\hline
\end{tabular}
}                                                              
\end{table*}

\begin{table*}
\addtocounter{table}{-1}
\caption{\small Continuation}
\label{macrotab}%
\centering{\small %
\begin{tabular}{llcclr@{}r@{}rr@{}r@{}rr}
\hline\hline
Num& Name                        &RA&Dec& SpT         & \multicolumn{3}{c}{MgI(\AA)} & \multicolumn{3}{c}{sTiO}& [Mg/Fe] \\                              
   &               &(h:m:s)  &(d:m:s)   &             & $\cal I$\ \ \ &$\sigma_{\rm typ}\cal I$\ &$\Delta\cal I$\ \ \ & $\cal I$\ \ \ &$\sigma_{\rm typ}\cal I$\ &$\Delta\cal I$\ \ \ \\                              
\hline
656&HD\,218658                         &23:07:53.9&$+$75:23:15&G2 III       &    0.660&\ \ 0.049&\    0.054& 0.958&\ \ 0.004&\    0.015&               \\
657&HD\,218857                         &23:11:24.6&$-$16:15:04&G6w          &    0.147&\ \ 0.112&\    0.014& 0.934&\ \ 0.009&\    0.004&   +0.15 $^{3}$\\
658&HD\,219134                         &23:13:17.0&$+$57:10:06&K3 Vvar      &    1.070&\ \ 0.044&\    0.047& 0.925&\ \ 0.004&\ $-$0.007& $-$0.016$^{1}$\\
659&HD\,219449                         &23:15:53.5&$-$09:05:16&K0 III       &    0.817&\ \ 0.036&\    0.056& 0.941&\ \ 0.003&\ $-$0.019&               \\ \medskip
660&HD\,219615                         &23:17:09.9&$+$03:16:56&G9 III       &    0.564&\ \ 0.060&\    0.000& 0.929&\ \ 0.005&\ $-$0.019&               \\
661&HD\,219617                         &23:17:05.0&$-$13:51:04&F8w          &    0.125&\ \ 0.071&\ $-$0.024& 0.922&\ \ 0.006&\    0.000&   +0.280$^{1}$\\
\hline
\end{tabular}
}                                                              
\end{table*}